\documentclass[journal]{IEEEtran} 

\usepackage{cite}
\usepackage{algorithm,algorithmicx}
\usepackage{algpseudocode}
\usepackage[small,bf]{caption}
\usepackage[cmex10]{amsmath}
\usepackage{amssymb,latexsym,epsfig,subfigure,epic,amscd,mathrsfs,euscript,eufrak,bm}
\usepackage{color}
\usepackage{booktabs}
\usepackage{array}
\usepackage{stfloats}
\usepackage{amsfonts}
\usepackage{amsopn}
\usepackage{graphicx}
\usepackage{url} 
\usepackage{empheq}
\usepackage{epstopdf}

\newtheorem{myprop}{\bf{Proposition}}

\newtheorem{remark}{\bf{Remark}}

\newcommand{\argmin}{\operatornamewithlimits{arg\,min}}

\DeclareMathOperator{\tr}{tr}

\DeclareMathOperator{\cov}{cov}
\DeclareMathOperator{\diag}{diag}
\DeclareMathOperator*{\minimize}{\text{minimize}}

\DeclareMathOperator*{\st}{\text{subject to}}
\DeclareMathAlphabet\mathbfcal{OMS}{cmsy}{b}{n}
\newcommand{\Def}[0]{\mathrel{\mathop:}=}

\begin{document}
 
\title{{Optimized Sensor Collaboration for Estimation of Temporally Correlated Parameters}}

\author{Sijia~Liu~\IEEEmembership{Member,~IEEE,}
        Swarnendu~Kar,~\IEEEmembership{Member,~IEEE,}
		Makan~Fardad,~\IEEEmembership{Member,~IEEE,}
        and~Pramod~K.~Varshney~\IEEEmembership{Fellow,~IEEE}
\thanks{Copyright (c) 2015 IEEE. Personal use of this material is permitted. However, permission to use this material for any other purposes must be obtained from the IEEE by sending a request to pubs-permissions@ieee.org.}   
\thanks{S. Liu was with Syracuse University. Now he is with the Department
of Electrical Engineering and Computer Science, University of Michigan, Ann Arbor, MI 48109, USA. Email:
lsjxjtu@umich.edu.}   
\thanks{S. Kar is with New Devices Group, Intel Corporation, Hillsboro, Oregon,
97124 USA. Email: swarnendu.kar@intel.com.}  
\thanks{M. Fardad and P. K. Varshney are with the Department
of Electrical Engineering and Computer Science, Syracuse University, Syracuse,
NY, 13244 USA. Email: \{makan, varshney\}@syr.edu.}
\thanks{The work of S. Liu and P. K. Varshney was supported by the U.S. Air Force
Office of Scientific Research  under grants FA9550-10-1-0458. The work of M. Fardad was supported by the National Science Foundation under awards EAGER ECCS-1545270 and CNS-1329885. }
}

\maketitle

\begin{abstract}
In this paper, {we aim to design the optimal sensor collaboration strategy 
 for the estimation of  time-varying parameters, where 
collaboration  refers to the act of sharing measurements with neighboring sensors prior to transmission to a fusion center.
We begin by addressing  the  sensor collaboration problem for the estimation of uncorrelated parameters. We show that the resulting collaboration problem can be transformed into a special nonconvex optimization problem, where a difference of convex functions carries all the nonconvexity. This specific problem structure enables the use of a convex-concave procedure to obtain a near-optimal solution.
When the parameters of interest are temporally correlated,  a penalized version of   the convex-concave procedure becomes well suited  for  designing the  optimal  collaboration scheme. 
In order to improve  computational efficiency, we  further propose a fast algorithm  that scales gracefully with problem size  via the alternating direction method of multipliers. 
Numerical results   are    
provided to demonstrate the effectiveness of our approach and the impact
of parameter correlation and temporal dynamics of   sensor networks on  estimation performance.}
\end{abstract}


\section{Introduction}

Wireless sensor networks (WSNs) consist of a large number of spatially distributed  sensors that often cooperate to perform parameter estimation; example applications  include environment monitoring, source localization and target tracking \cite{olirod11,zoucha04,hevicyan06}. 
Under limited resources, such as limited communication bandwidth and sensor battery power, it is important to design an energy-efficient architecture for distributed estimation.
In this paper, we employ a WSN to estimate time-varying parameters in the presence of inter-sensor communication that  is referred to as  {sensor collaboration}. Here sensors 
are  allowed to update their measurements by taking a linear combination of the measurements of those they interact with
 prior to transmission to a fusion center (FC). 
The presence of sensor collaboration  smooths out
the  observation  noise,  thereby  improving  the  quality  of  the
signal and the eventual estimation performance. 



Early research efforts \cite{cuixiagolluopoo07,xiacuiluogol08,marmattong07,sargas10,liusay07,mernawton07,bansmi10,cohles13,mayveggal15,dastep14} focused on the problem of distributed inference (estimation or detection) in the absence of sensor collaboration, where an  
 amplify-and-forward transmission strategy is commonly used.
In \cite{cuixiagolluopoo07}, the problem of designing optimal power amplifying factors (also known as power allocation problem) was studied for distributed estimation  over an orthogonal multiple access channel (MAC). 
{In \cite{xiacuiluogol08}, the power allocation problem   was addressed  when the MAC is coherent, where sensors coherently  form  a  beam  into  a  common  channel  received at the FC. In \cite{marmattong07}, a likelihood-based multiple access communication strategy was proposed for estimation, and was proved to be asymptotically efficient as the number of sensors increases. In \cite{sargas10},  feedback signals   were studied to combat uncertainty in the observation model for distributed estimation with coherent MAC. 
In \cite{liusay07}, distributed detection problem was studied in the setting of identical Gaussian multiple access  channels (without fading). It was shown that  the centralized error exponent can be achieved via the transmission of the log-likelihood ratio as the number of sensors  approaches infinity.
Further in \cite{mernawton07,bansmi10,cohles13}, asymptotic detection performance was  studied over multiaccess fading channels. In \cite{mayveggal15}, the problem of power allocation was studied for distributed detection using a MAC.
In \cite{dastep14}, the impact of nonlinear bounded transmission schemes was studied on distributed detection and estimation. In the aforementioned literature   \cite{cuixiagolluopoo07,xiacuiluogol08,marmattong07,sargas10,liusay07,mernawton07,bansmi10,cohles13,mayveggal15,dastep14},
the act of   inter-sensor communication was not considered. In contrast, here we seek the optimal sensor collaboration scheme for  the estimation of temporally correlated parameters.}


Recently, the problem of distributed estimation with sensor
collaboration has attracted  attention \cite{fanli09,thamit06_asilomar, thamit08,fanvaljamsch_14,karvar13,karvar12isit,karvar12allerton,liukarfarvar14_isit,liuswafarvar15}.
In  \cite{fanli09}, the optimal power allocation strategy was found for a fully connected network, where all the sensors are allowed to collaborate, namely, share their measurements with the other sensors.
It was shown that sensor collaboration results in significant
improvement of estimation performance compared with the
conventional amplify-and-forward transmission scheme.
In \cite{thamit06_asilomar} and \cite{thamit08},
  optimal power allocation schemes
were found for star, branch and linear network topologies.
In \cite{fanvaljamsch_14}, the sensor collaboration problem was studied for parameter   estimation via the best linear unbiased estimator. 
In \cite{karvar13,karvar12isit,karvar12allerton}, the problem of sensor collaboration was studied given an arbitrary   collaboration topology. It was observed that even a partially connected network can yield performance close to that of a fully connected network.
In \cite{liukarfarvar14_isit} and \cite{liuswafarvar15},  nonzero collaboration costs were taken into account, and a sparsity inducing optimization
framework was proposed to jointly design both sensor selection and sensor collaboration schemes. 

{In the existing  literature \cite{fanli09,thamit06_asilomar, thamit08,fanvaljamsch_14,karvar13,karvar12isit,karvar12allerton,liukarfarvar14_isit,liuswafarvar15}, 
 sensor collaboration was studied in {static} networks, where sensors take a {single snapshot} of the static parameter, and then initiate sensor collaboration protocols designed in the setting of single-snapshot estimation.
In contrast, here we study the problem of sensor collaboration for the estimation of \textit{temporally-correlated} parameters in \textit{dynamic} networks that involve, for example, time-varying observation and channel gains.}
{Solving such a problem is also motivated by real-life applications, in which the physical phenomenon
to be monitored such as daily temperature, precipitation, soil moisture  and  seismic activities \cite{vurakaaky04,vuraka06,P2004}  is   temporally correlated. For example, when monitoring  daily temperature variations,  temperatures at different times of the day are strongly correlated, e.g., a cold morning is likely to be followed by a cold afternoon.}

Due to the presence of temporal dynamics and parameter correlation,  
 optimal   sensor collaboration schemes at
multiple time steps  are coupled with each other, and thus
pose many challenges in problem formulation and optimization
compared to the existing work  \cite{fanli09,thamit06_asilomar, thamit08,fanvaljamsch_14,karvar13,karvar12isit,karvar12allerton,liukarfarvar14_isit,liuswafarvar15}.
For example, when parameters of interest are temporally correlated,  
expressing the estimation distortion in a succinct closed form (with respect to the   collaboration variables) is not straightforward. 
It should be pointed out that even  for uncorrelated parameters, finding the optimal collaboration scheme for each time step is  nontrivial since energy constraints are temporally inseparable. 
In this paper, we seek the optimal sensor collaboration scheme  by minimizing the   estimation distortion   subject to individual energy constraints of sensors in the presence of (a) temporal dynamics in system, (b) temporal correlation of parameter, and (c) energy constraints in time.

{
Besides  \cite{fanli09,thamit06_asilomar, thamit08,fanvaljamsch_14,karvar13,karvar12isit,karvar12allerton,liukarfarvar14_isit,liuswafarvar15}, 
our work is also related to but quite different from the problem of consensus-based decentralized estimation \cite{karmou09,hosschapmes14,sharak16,olf07,carchisch08,schrib08,schgiarou08}. The common idea in \cite{karmou09,hosschapmes14,sharak16,olf07,carchisch08,schrib08,schgiarou08} is that the task of centralized estimation can be performed  using local estimators at sensors together with inter-sensor communications. It was shown in  \cite{schrib08} and \cite{schgiarou08} that the success of  decentralized estimation is based on the fact that the {global} estimation cost with respect to the parameter of interest   can be converted into a {sum} of {local} cost functions subject to consensus constraints. {Different from  \cite{karmou09,hosschapmes14,sharak16,olf07,carchisch08,schrib08,schgiarou08}, 
the focus of this paper is to
 design the optimal energy allocation strategy (namely, the collaboration weights),  rather  to find the
optimal estimate.} Here  tasks of estimation and optimization are completed at an FC. Moreover, the studied sensor network is  {not} necessarily  {connected}.  
An extreme case is that in the absence of inter-sensor communication, the proposed sensor collaboration problem would reduce to the conventional power allocation problem (based on the amplify-and-forward transmission strategy)  \cite{xiacuiluogol08,cuixiagolluopoo07}. Therefore, our problem is different from the  consensus-based decentralized estimation problem, in which the network is assumed to be connected so that the consensus of estimate at local sensors can be  achieved. 
}

In our work, design of the optimal   collaboration scheme  is studied  under two scenarios: a) parameters are temporally uncorrelated or  prior knowledge about temporal correlation is not available, and b) parameters are   temporally  correlated.  
When parameters  are   uncorrelated,  we derive the closed form of the estimation distortion with respect to   sensor collaboration variables, which  is in the form of  a sum  of quadratic ratios. 
We show that the resulting  sensor collaboration problem is equivalent to a nonconvex quadratically constrained  problem, in which the difference of convex    functions carries  all the nonconvexity.   This
specific problem structure enables the use of convex-concave procedure (CCP) \cite{yuiran03} to solve the sensor collaboration problem in a numerically efficient manner. 


When parameters of interest are temporally  correlated, expressing the estimation error as an explicit function of the collaboration variables becomes    difficult. In this case, we show that the  sensor collaboration problem can be converted into a semidefinite program   together with   a  (nonconvex) rank-one constraint. 
After convexification, 
  the method of penalty CCP \cite{lipboy14}  becomes well-suited for seeking the optimal sensor collaboration scheme.
However,  
the proposed algorithm
is computationally intensive for   large-scale problems.
To
improve computational efficiency, we    develop a fast algorithm that scales gracefully with problem size by using the alternating direction method of multipliers (ADMM) \cite{boyparchupeleck11}. 

We summarize our contributions as follows.
\begin{itemize}
\item {We propose a tractable optimization framework for the
  design of the optimal   collaboration scheme that accounts for    parameter correlation and temporal dynamics of sensor networks}.
\item We show that the  problem of sensor collaboration   for the estimation of temporally uncorrelated parameters can  be solved as a special nonconvex   problem, where   the only source of nonconvexity  can be isolated to a constraint that contains the difference of convex functions.
\item We provide valuable insights into the problem structure of sensor collaboration  with      correlated parameters,  and propose an ADMM-based   algorithm for improving  the computational efficiency.
\end{itemize}


The rest of the paper is organized as follows. In Section\,\ref{sec: sys}, we introduce the collaborative estimation system, and present the    general formulation of   the optimal sensor collaboration problem. In Section\,\ref{sec: simplify}, we discuss two types of sensor collaboration problems for the estimation of temporally uncorrelated and correlated parameters. In Section\,\ref{sec: col_uncorr}, we study the sensor collaboration problem with uncorrelated parameters.  In Section\,\ref{sec: LMMSE_gen}, we propose efficient optimization methods to  solve the sensor collaboration problem with correlated parameters.
 In Section\,\ref{sec: numerical}, we  demonstrate the effectiveness of our       approach  through numerical examples. Finally, in
Section\,\ref{sec: conc} we summarize our work and discuss   future research directions.


\section{System Model}
\label{sec: sys}
\begin{figure}[htb]
\centering
\includegraphics[width=.48\textwidth,height=!]{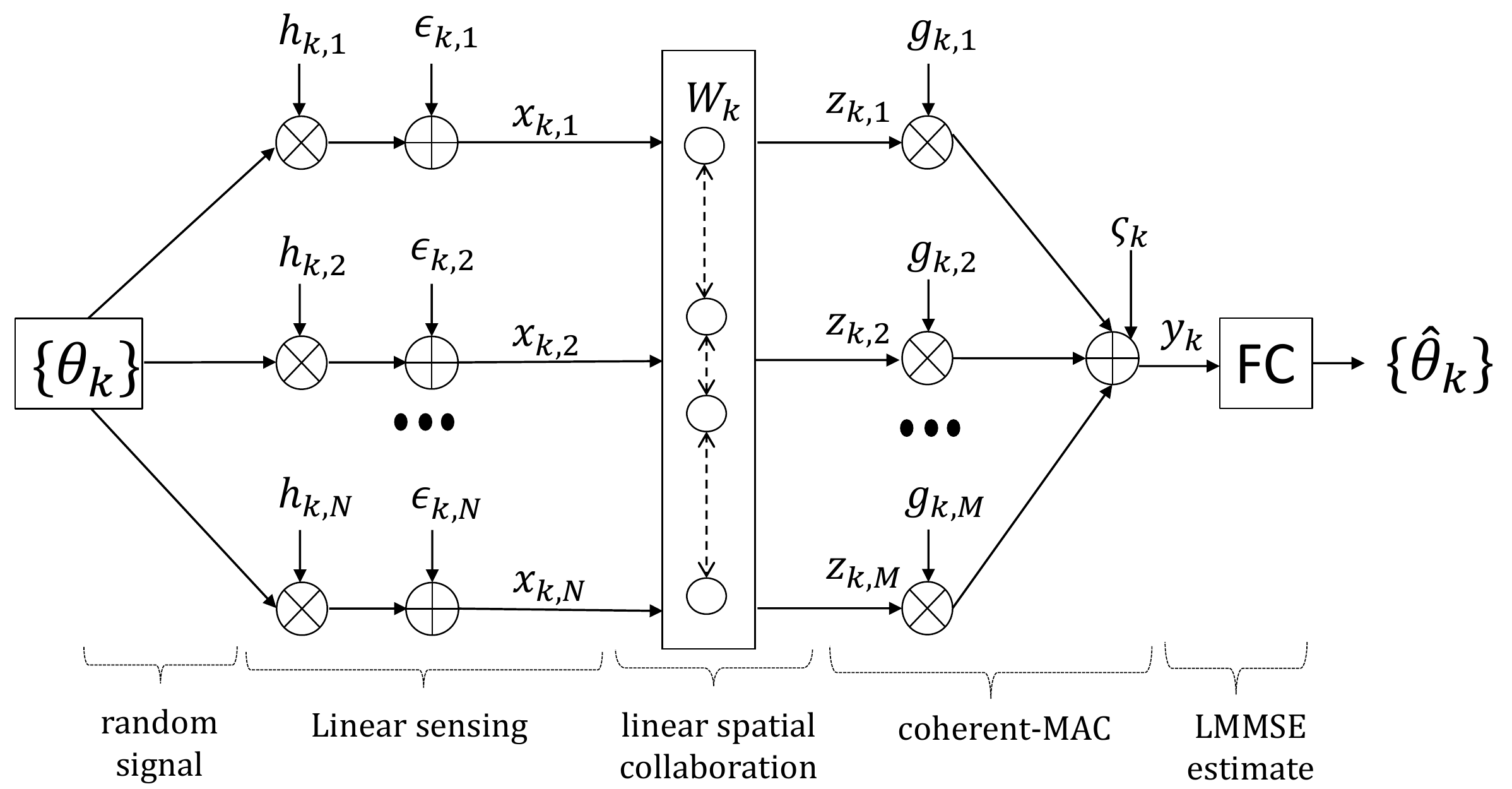} 
\caption{\footnotesize{Collaborative estimation architecture.
}}
  \label{fig: sysmodel}
\end{figure}

In this section, we introduce the collaborative  estimation system and
 formulate  the sensor collaboration problem considered in this work. 
The task here is to estimate a   time-varying parameter $\theta_k$ over a time horizon of length $K$. In the estimation system, sensors first accquire their raw measurements via a linear sensing model, and then
update their observations through spatial collaboration, where collaboration refers to the act of sharing measurements
with neighboring sensors. The collaborative signals are then transmitted through a coherent MAC to the FC, which finally determines a global estimate of $\theta_k$ for $k \in [K]$.
The overall architecture of the   collaborative estimation system is shown in Fig.\,\ref{fig: sysmodel}.

The vector of measurements from $N$ sensors at time   $k$ is given by the linear sensing model
\begin{align}
\mathbf x_k =  {\mathbf h_k} \theta_k + \boldsymbol \epsilon_k, ~ k \in [K],
\label{eq: lin_sen}
\end{align}
where for notational simplicity, let $[K]$ denote the integer set $\{1,2,\ldots,K \}$,
$\mathbf x_k = [x_{k,1}, \ldots,x_{k,N}]^T$ is the vector of measurements, $\mathbf h_k = [h_{k,1},  \ldots,h_{k,N}]^T$ is the vector of observation gains,
without loss of generality $\theta_k$ is assumed to be  a   random   process with   zero mean and   variance $\sigma_{\theta}^2$, 
 $\boldsymbol \epsilon_k = [\epsilon_{k,1}, \ldots, \epsilon_{k,N}]^T$ is the vector of Gaussian noises with i.i.d variables $\epsilon_{k,n} \sim \mathcal N(0, \sigma_{\epsilon}^2)$ for $k \in [K]$ and $n \in [N]$.

After linear sensing, each sensor may pass its observation
to other sensors for collaboration prior to transmission to the
FC. With a relabelling of   sensors, we assume that the first $M$ sensors (out of a total of $N$ sensor nodes) communicate with the FC. Collaboration   among     sensors is represented  by a  {known} matrix $  {\mathbf A} \in \mathbb R^{M \times N}$ with zero-one entries, namely, $A_{mn} \in \{ 0, 1\}$ for $m \in [M]$ and $n \in [N]$. {Here   we call $\mathbf A$ a topology matrix, where 
 $A_{mn}  = 1$ signifies that    the $n$th sensor shares its observation with the $m$th sensor, and $A_{mn}  = 0$ indicates the absence of a collaboration link from the $n$th sensor to the $m$th sensor. Note that $\mathbf A$ is
essentially a truncated adjacency matrix.} 
{The bidirectional communication link between two sensors indicates that the underlying graph of the network is directed but not necessarily connected. In particular, the network given by $A_{mn} = 0$ for $n \neq m$ corresponds to the amplify-and-forward transmission strategy considered in \cite{cuixiagolluopoo07}.} 

{
Based on the topology matrix,   the sensor collaboration process  at time $k$ is given by
\begin{align}
\begin{array}{l}
\mathbf z_k =  \mathbf W_k \mathbf x_k,   ~k \in [K] \vspace*{0.03in} \\
\mathbf W_k \circ (\mathbf 1_M \mathbf 1_N^T - \mathbf A) = \mathbf 0, 
\end{array}
\label{eq: col_sig}
\end{align}
where  $\mathbf z_k= [z_{k,1}, z_{k,2}, \ldots, z_{k,M}]^T$,  $z_{k,m}$ is  the    signal after collaboration at  sensor $m$ and time $k$,      $\mathbf W_k \in \mathbb R^{M \times N}$ is the collaboration matrix that contains collaboration weights (based on the energy allocated) used to combine sensor measurements   at time $k$, $\circ$ denotes the elementwise product,  $\mathbf 1_N$ is the $N \times 1$ vector of all ones,   and $\mathbf 0$ is the $M \times N$ matrix of all zeros. 
In what follows, 
while refering to vectors of all ones and all zeros, their   dimensions will be omitted for simplicity  but can be inferred from the context.
In \eqref{eq: col_sig}, we assume that sharing of an observation is realized through an ideal (noise-less and cost-free) communication link. The proposed ideal  collaboration model enables us to  obtain  explicit expressions for   transmission cost and   estimation
distortion. 
}

After sensor collaboration, the message $\mathbf z_k$
is transmitted through a coherent
MAC so that the received signal $y_k$ at the FC is a coherent sum \cite{xiacuiluogol08} 
\begin{align}
y_k = \mathbf g_k^T \mathbf z_k + \varsigma_k, ~ k \in [K],
\label{eq: channel}
\end{align}
where 
${\mathbf g_k = [g_{k,1,},g_{k,2},\ldots,g_{k,M}]^T}$ is the vector of channel
gains,  and  $ \varsigma_k $ is temporally independent Gaussian noise with zero mean and   variance $\sigma_{\varsigma}^2$.



From  \eqref{eq: lin_sen}\,--\,\eqref{eq: channel}, the vector of  received signals at the FC   can be compactly expressed as a linear function of  parameters 
$\boldsymbol \theta = [\theta_1, \theta_2, \ldots, \theta_K]^T$,
\begin{align}
\mathbf y = \mathbf D_W \mathbf D_h \boldsymbol \theta + \boldsymbol \nu, ~ \mathbf D_W \Def \mathrm{blkdiag} \{   \mathbf g_k^T \mathbf W_k  \}_{k=1}^K,
\label{eq: y_theta}
\end{align}
where  $\mathbf y = [y_1, y_2, \ldots, y_K]^T$, $\boldsymbol \nu  = [\nu_1, \nu_2, \ldots, \nu_K]^T$, $
\nu_k \Def  \mathbf g_k^T \mathbf W_k \boldsymbol \epsilon_k + \varsigma_k
$, 
   $\mathbf D_h \Def \mathrm{blkdiag}\{ \mathbf h_k\}_{k=1}^K$, and 
$\mathrm{blkdiag}\{\mathbf X_i\}_{i=1}^n$ denotes the block-diagonal matrix with diagonal blocks $\mathbf X_1, \mathbf X_2, \ldots, \mathbf  X_n$.

At the FC, we employ a linear minimum mean squared-error estimator  (LMMSE) \cite{karbook} to  estimate $\boldsymbol \theta$, where we assume that the FC knows  the observation gains, channel gains, and the
second-order statistics of   the parameters of interest and additive
noises. 
The corresponding 
estimation error covariance   is given by \cite[Theorem\,10.3]{karbook}
\begin{align}
\mathbf P_W = ( \boldsymbol \Sigma_{\theta}^{-1} + \mathbf D_h^T \mathbf D_W^T \mathbf D_\nu^{-1} \mathbf D_W \mathbf D_h )^{-1},
\label{eq: MSE_LMMSE}
\end{align}
where $\boldsymbol \Sigma_{\theta}$ represents  prior knowledge about the parameter correlation,  particularly  $\boldsymbol \Sigma_{\theta} = \sigma_\theta^2 \mathbf I_K$ for temporally uncorrelated parameters, $\mathbf I_K$ is the $K \times K$ identity matrix, and
$\mathbf D_{\nu}  \Def \sigma_{\epsilon}^2 \mathbf D_W \mathbf D_W^T +\sigma_{\varsigma}^2 \mathbf I_K$.
It is clear from \eqref{eq: MSE_LMMSE} that  the estimation error covariance matrix is a function of collaboration matrices $\{ \mathbf W_k \}$, and their dependence  on $\{ \mathbf W_k \}$ is through $\mathbf D_{W}$. This  dependency  does  not  lend  itself  to  easy optimization of scalar-valued functions of  $\mathbf P_W$ for design of the optimal sensor collaboration scheme. 
More insights into the  LMMSE  will be provided in Sec.\,\ref{sec: simplify}.


 We next define the transmission cost of the $m$th sensor at time $k$, which refers
to the energy consumption of transmitting the collaborative
message $\mathbf z_k$ to the FC. That is,
\begin{align}
 T_{m}(\mathbf W_k)   & =   \mathbb E_{\theta_k, \boldsymbol \epsilon_k } [z_{k,m}^2]
 \nonumber \\
& =       \mathbf e_m^T \mathbf W_k (\sigma_{\theta}^2\mathbf h_k\mathbf h_k^T +  \sigma_{\epsilon}^2 \mathbf I_N)  \mathbf W_k^T \mathbf e_m,
\label{eq: P_W}
\end{align}
for $m \in [M]$ and $k \in [K]$, where $\mathbf e_m \in \mathbb R^M$ is   a basis vector
with $1$ at the $m$th coordinate and $0$s elsewhere.
In what follows, 
while refering to basis vectors and identity matrices, their   dimensions will be omitted for simplicity  but can be inferred from the context.

We now state the main optimization problem considered in this work for sensor collaboration 
\begin{align}
\begin{array}{llr}
\displaystyle \minimize &  \displaystyle \tr \left (  \mathbf P_W \right )  & \vspace*{0.03in} \\
\st 
&  \displaystyle \sum_{k = 1}^K T_{m} (\mathbf W_k)  \leq E_{m},  & m \in [M] \vspace*{0.03in}\\
&  \mathbf W_k \circ (\mathbf 1_M \mathbf 1_N^T - \mathbf A) = \mathbf 0,  & k \in [K],
\end{array}
\label{eq: prob_main}
\end{align}
where  
$\mathbf W_k $ is the optimization variable  for  $k \in [K]$,
  $\tr(\mathbf P_W)$ denotes the estimation distortion of using the LMMSE, 
$T_{m} (\mathbf W_k) $ is the transmission cost given by \eqref{eq: P_W},  $E_m$ is   a prescribed   energy budget of   the $m$th sensor,  and ${\mathbf A}$  characterizes the network topology.
The problem structure and the solution of 
   \eqref{eq: prob_main} will be elaborated on in the rest of the paper.

{We end this section with the following remarks.
\begin{remark}
In the system model, the assumption of known observation and channel
gains can be further relaxed to that of given knowledge about
their second-order statistics. Our earlier work \cite{liuswafarvar15} has shown
that under this weaker assumption, we can obtain similar
expressions of the linear estimator. In this paper, we assume
the observation and channel models are known
for ease of presentation and analysis.
\end{remark}
\begin{remark}
{Although sensor collaboration is performed with respect to a time-invariant (fixed) topology matrix  ${\mathbf A}$,   energy allocation in terms of  the magnitude of nonzero entries in $\mathbf W_k$ is time varying in the presence of  temporal dynamics of the sensor network.} 
As will be evident later, 
the proposed sensor collaboration approach is also applicable to   the problem with time-varying topologies. 
\end{remark}
}

\section{Reformulation and Simplification Using Matrix Vectorization}
\label{sec: simplify}
In this section, we  simplify
problem \eqref{eq: prob_main}  by 
exploiting the sparsity structure of the topology matrix  and concatenating the  nonzero  entries of a collaboration matrix into a collaboration vector.  
{There exist two  benefits to using   matrix vectorization:  
 a)  the topology constraint  in \eqref{eq: prob_main} can be eliminated without loss of performance, which  renders a less complex problem; b)  the structure of  nonconvexities  is more easily revealed via such a  reformulation. 
}

 \begin{figure}[htb]
\centering
\includegraphics[width=.465\textwidth,height=!]{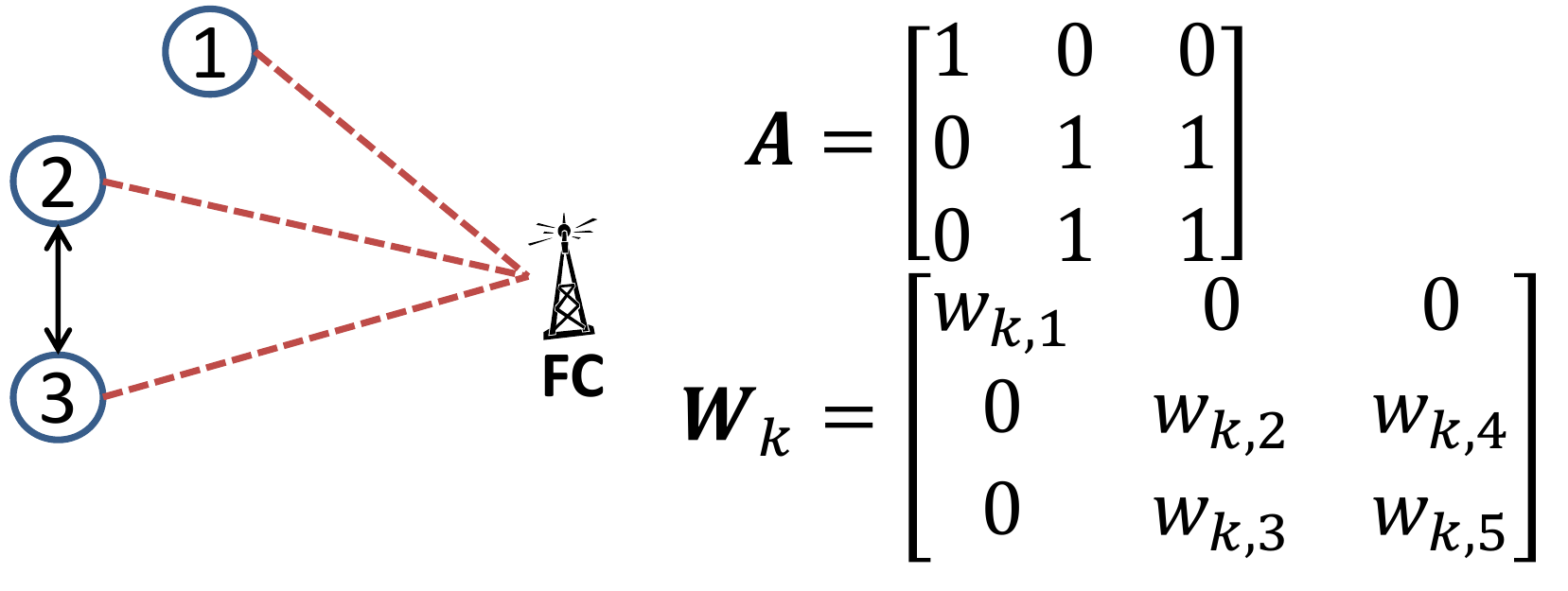} 
\caption{\footnotesize{Example of vectorization of $\mathbf W_k$.
}}
  \label{fig: vectorization}
\end{figure}

In problem \eqref{eq: prob_main}, the only optimization variables  are the nonzero entries of   collaboration matrices. We concatenate these nonzero entries (columnwise)     into a collaboration vector
\begin{align}
\mathbf w_k = [w_{k,1}, w_{k,2}, \ldots,w_{k,L}]^T, 
\label{eq: colW}
\end{align}
where  $w_{k,l}$ denotes the $l$th entry of $\mathbf w_k$, and $L$ is the number of nonzero entries of the topology matrix ${\mathbf A}$.
We note that given $  w_{k,l}$, there exists a row index $m_l$ and a column index $n_l$ such that $w_{k,l} = [\mathbf W_k]_{m_l n_l}$, where
$[\mathbf X]_{mn}$ (or $X_{mn}$) denotes the $(m,n)$th entry of a matrix $\mathbf X$.
 We demonstrate the vectorization of $\mathbf W_k$ through an example in Fig.\,\ref{fig: vectorization}, where we consider $N = 3$ sensor nodes, $M=3$ communicating nodes, and $2$     collaboration links.

\subsection{Collaboration problem for the estimation of  uncorrelated   parameters}

When the  parameters of   interest are  uncorrelated,   
the estimation error covariance  matrix   \eqref{eq: MSE_LMMSE}
simplifies to  
\begin{align}
&   \mathbf P_W  =   \left (  \sigma_{\theta}^{-2} \mathbf I  + \mathbf D_h^T \mathbf D_W^T (\sigma_{\epsilon}^2\mathbf D_W \mathbf D_W^T +\sigma_{\varsigma}^2 \mathbf I )^{-1} \mathbf D_W \mathbf D_h \right )^{-1}  \nonumber \\
 =&   \left (  \sigma_{\theta}^{-2} \mathbf I  +    \diag \left \{ \frac{ \mathbf g_k^T \mathbf W_k \mathbf h_k \mathbf h_k^T \mathbf W_k^T \mathbf g_k}{\sigma_{\epsilon}^2 \mathbf g_k^T \mathbf W_k \mathbf W_k^T \mathbf g_k+\sigma_{\varsigma}^2 } \right \}_{k=1}^K   \right)^{-1} \nonumber \\
  = &
    \diag \hspace*{-0.03in} \left \{ \frac{ \sigma_\theta^{2} \sigma_{\epsilon}^2    \mathbf g_k^T \mathbf W_k   \mathbf W_k^T \mathbf g_k  +\sigma_\theta^{2} \sigma_{\varsigma}^2 }{\sigma_\theta^{2} \mathbf g_k^T \mathbf W_k \mathbf h_k \mathbf h_k^T \mathbf W_k^T \mathbf g_k  \hspace*{-0.03in}  +  \hspace*{-0.03in}   \sigma_{\epsilon}^2 \mathbf g_k^T \mathbf W_k \mathbf W_k^T \mathbf g_k \hspace*{-0.03in}  + \hspace*{-0.03in}  \sigma_{\varsigma}^2  } \right \}_{k=1}^K,  
\label{eq: MSE_IID_Close}
\end{align}
where  
$\diag\{a_k\}_{k=1}^K$ denotes a diagonal matrix with diagonal entries $a_1$, $a_2$, \ldots, $a_K$.



Let $\mathbf w \in \mathbb R^L$ be the vector obtained by stacking the nonzero entries of $\mathbf W \in \mathbb R^{M \times N}$ columnwise. Then
\begin{align}
\mathbf b^T \mathbf W = \mathbf w^T \mathbf B, \label{eq: prop1}
\end{align}
where $\mathbf b \in \mathbb R^{N}$ is a   coefficient vector,    $\mathbf B $ is an $L \times N$ matrix whose $(l,n)$th entry is given by
\begin{align}
B_{ln} = \left \{
\begin{array}{ll}
b_{m_l} &  n = n_l \\
0 & \text{otherwise},
\end{array}
\right. 
\label{eq: A_wW}
\end{align}
and
 the indices $m_l$ and $n_l$ are such that $w_l = W_{m_l n_l}$ for $l \in [L]$.
The proof of equation \eqref{eq: prop1} is given in  Appendix\,\ref{appendix: Prop1} for the sake of completeness.


From \eqref{eq: MSE_IID_Close}  and \eqref{eq: prop1}, 
the objective function of  problem \eqref{eq: prob_main} can be rewritten as
\begin{align}
\phi(\mathbf w) \Def \tr(\mathbf P_W) =   \sum_{k=1}^K \frac{ \sigma_\theta^{2} \sigma_{\epsilon}^2    \mathbf w_k^T  \mathbf R_k \mathbf w_k  +\sigma_\theta^{2} \sigma_{\varsigma}^2 }{\mathbf w_k^T \mathbf S_k \mathbf w_k    +    \sigma_{\varsigma}^2  }  ,  \label{eq: Pw}
\end{align}
where we used the fact  that  $\mathbf g_k^T \mathbf W_k = \mathbf w_k^T \mathbf G_k$,  i.e., $\mathbf G_k$ is derived
  from $\mathbf g_k$ in the same way that $\mathbf B$ is derived from $\mathbf b$ in \eqref{eq: prop1},
and $\mathbf S_k \Def \mathbf G_k (\sigma_\theta^2 \mathbf h_k \mathbf h_k^T + \sigma_\epsilon^2 \mathbf I) \mathbf G_k^T$. 

Moreover, the transmission cost   \eqref{eq: P_W} can be  rewritten as
\begin{align}
& T_{m}(\mathbf w_k)   \Def         \mathbf w_k^T \mathbf Q_{k,m}  \mathbf w_k, \label{eq: Tw} \\ 
&\mathbf Q_{k,m} \Def \mathbf E_m  (\sigma_{\theta,k}^2\mathbf h_k\mathbf h_k^T +  \sigma_{\epsilon}^2 \mathbf I )\mathbf E_m^T, \nonumber
\end{align}
where 
$\mathbf E_m$ is defined as in \eqref{eq: prop1} such that $\mathbf e_m^T \mathbf W_k = \mathbf w_k^T \mathbf E_m$.  We remark that $ \mathbf Q_{k,m}$ is   positive semidefinite   for $k \in [K]$ and $m \in [M]$. 

From \eqref{eq: Pw} and \eqref{eq: Tw}, the sensor collaboration problem for the estimation of temporally uncorrelated   parameters becomes
\begin{align}
\begin{array}{ll}
 \minimize &  \phi(\mathbf w)
\vspace*{0.03in} \\
\st 
&  
\mathbf w^T \mathbf Q_{m}  \mathbf w  \leq E_{m},  ~    m \in [M], 
\end{array}
\label{eq: prob_main_uncorr}
\tag{P1}
\end{align}
where $\mathbf w = [\mathbf w_1^T, \mathbf w_2^T, \ldots, \mathbf w_K^T]^T$ is the optimization variable,   $\phi(\mathbf w)$ is the estimation distortion given by \eqref{eq: Pw}, and $\mathbf Q_m \Def \mathrm{blkdiag}\{ \mathbf Q_{k,m }\}_{k=1}^K$.
Note that \eqref{eq: prob_main_uncorr} cannot be decomposed  in time since   sensor energy constraints are temporally inseparable.  

Compared to problem \eqref{eq: prob_main}, 
 the topology constraint  in terms of $\mathbf A$ is eliminated without loss of performance  in \eqref{eq: prob_main_uncorr} since  the sparsity  structure of the topology matrix has  been taken into account while constructing the   collaboration vector. 
{In the special case of single-snapshot estimation (namely, $K = 1$),
 the objective function of \eqref{eq: prob_main_uncorr}  simplifies to a single quadratic ratio. It has been shown in   \cite{karvar13} and \cite{liuswafarvar15} that such a nonconvex problem   can be readily solved  via convex programming. In contrast,  \eqref{eq: prob_main_uncorr} is a more complex nonconvex optimization problem, where the nonconvexity stems from  the \textit{sum} of quadratic ratios in the objective function.  As indicated in  \cite{shenchen09} and \cite{bughenlas15},  
the Karush-Kuhn-Tucker (KKT) conditions of such a complex fractional optimization problem are intractable to solve to obtain the globally optimal solution (or all  locally optimal solutions). Therefore,   
an efficient local optimization method will be proposed to solve   
 \eqref{eq: prob_main_uncorr}  in Sec.\,\ref{sec: col_uncorr}.   Also, the efficacy of the proposed solution will be  shown in Sec.\,\ref{sec: numerical} via extensive numerical experiments.
}

\subsection{Collaboration problem for the estimation of   correlated   parameters}


When parameters are temporally correlated, the covariance matrix $\boldsymbol \Sigma_{\theta}$ is no longer diagonal and it is not straight forward to express the estimation error in a succinct form, as was done in \eqref{eq: Pw}.
We recall  from \eqref{eq: MSE_LMMSE}   that 
the dependence of the estimation error covariance  on collaboration matrices is through $\mathbf D_W$. According to the  matrix inversion lemma \cite[A\,1.1.3]{karbook}, 
\begin{align}
  & 
\mathbf D_W^T (\sigma_{\epsilon}^{2}\mathbf D_W  \mathbf D_W^T + \sigma_{\varsigma}^{2} \mathbf I)^{-1} \mathbf D_w
   \nonumber \\
& \hspace*{0.6in}=     \sigma_{\epsilon}^{-2}\mathbf I  - (\sigma_{\epsilon}^2 \mathbf I + \sigma_{\epsilon}^4 \sigma_{\varsigma}^{-2} \mathbf D_W^T \mathbf D_W)^{-1}.
\label{eq: matrix_inversion}
\end{align}
Substituting  \eqref{eq: matrix_inversion} into \eqref{eq: MSE_LMMSE}, we obtain
\begin{align}
 \mathbf P_{W} 
= 
\left ( \mathbf C - \sigma_{\epsilon}^{-2} \mathbf D_h^T
( \mathbf I + \sigma_{\epsilon}^{2}  \sigma_{\varsigma}^{-2} \mathbf D_W^T \mathbf D_W)^{-1}  \mathbf D_h \right )^{-1}
\label{eq: MSE_LMMSE_v1}
\end{align}
with  $\mathbf C \Def \boldsymbol \Sigma_{\theta}^{-1} + \sigma_{\epsilon}^{-2} \mathbf D_h^T \mathbf D_h$. According to the definition of $\mathbf D_W$ in \eqref{eq: y_theta}, we obtain
\begin{align}
 \mathbf D_W^T \mathbf D_W &= \mathrm{blkdiag} \{ \mathbf W_k^T \mathbf g_k \mathbf g_k^T \mathbf W_k \}_{k=1}^K \nonumber \\
& = \mathrm{blkdiag} \{ \mathbf G_k^T \mathbf w_k \mathbf w_k^T \mathbf G_k \}_{k=1}^K,
\label{eq: DwDw}
\end{align}
 where $\mathbf G_k$ 
has been introduced in the paragraph that proceeds \eqref{eq: Pw}.

Combining   \eqref{eq: MSE_LMMSE_v1} and \eqref{eq: DwDw}, we can rewrite the estimation error covariance as a function of the collaboration vector
\begin{align}
& \mathbf P_w \Def    \left ( \mathbf C \hspace*{-0.02in} - \hspace*{-0.02in} \sigma_{\epsilon}^{-2} \diag  {\big \{} \mathbf h_k^T \left (  \mathbf I  \hspace*{-0.02in} + \hspace*{-0.02in} \sigma_{\epsilon}^{2}  \sigma_{\varsigma}^{-2} \mathbf G_k^T \mathbf w_k \mathbf w_k^T \mathbf G_k \right )^{-1}  \right.   \nonumber \\
&  \hspace*{1.27in} \left.  \,  \cdot \,  \mathbf h_k {\big  \}}_{k=1}^K \right )^{-1}.
\label{eq: MSE_LMMSE_final}
\end{align}

From \eqref{eq: MSE_LMMSE_final}, the sensor collaboration problem   for the estimation of  temporally correlated   parameters becomes
\begin{align}
\begin{array}{ll}
\displaystyle \minimize &~ \tr(\mathbf P_w) \vspace*{0.03in}\\
\st 
&~  
\mathbf w^T \mathbf Q_{m} \mathbf w \leq E_{m},  ~ m \in [M],
\end{array}
\label{eq: prob_main_LMMSE}
\tag{P2}
\end{align}
where $\mathbf w = [\mathbf w_1^T, \mathbf w_2^T, \ldots, \mathbf w_K^T]^T$ is the optimization variable. 

{We note that \eqref{eq: prob_main_LMMSE} is a nonconvex optimization problem. We will show in Sec.\,\ref{sec: LMMSE_gen} that the  rank-one matrix $\mathbf w_k \mathbf w_k^T$ that appears in \eqref{eq: MSE_LMMSE_final} is the source of nonconvexity. 
Compared to \eqref{eq: prob_main_uncorr},    \eqref{eq: prob_main_LMMSE}  is more involved due to the presence of the parameter correlation. 
We will also show  that  
\eqref{eq: prob_main_LMMSE} 
can be cast as a particular nonconvex optimization problem, where the objective function is linear, and the constraint set is formed by  convex quadratic constraints, linear matrix inequalities and nonconvex rank constraints.
The presence of generalized inequalities (with respect to positive semidefnite cones) and  rank constraints make    KKT conditions complex and intractable to find the globally optimal solution. Instead, we will employ an  efficient convexification method to find a locally optimal solution of \eqref{eq: prob_main_LMMSE}.   The efficacy of the proposed optimization method will be empirically shown in Sec.\,\ref{sec: numerical}.
}

{We finally remark that both  \eqref{eq: prob_main_uncorr} and \eqref{eq: prob_main_LMMSE} are   feasible optimization problems, namely,    in the sense that an optimal solution exists for each of them. 
This can be examined as follows. First, there
exists a non-empty constraint set. For example, $\mathbf w_k = \mathbf 0$ is a feasible solution to \eqref{eq: prob_main_uncorr} and \eqref{eq: prob_main_LMMSE}. When $\mathbf w_k = \mathbf 0$, the estimate of the unknown parameter is only
determined by the prior knowledge about the parameter. Second, the optimal
value is bounded due to the presence of the energy constraint.}

\section{Special Case: Optimal Sensor Collaboration for  The Estimation of  Uncorrelated   Parameters}
\label{sec: col_uncorr}
{In this section, we show  that \eqref{eq: prob_main_uncorr} can be transformed into a special nonconvex optimization problem, where the difference of convex (DC)  functions carries all the nonconvexity.} Spurred by the problem structure,
we  employ   a convex-concave procedure (CCP) to solve  \eqref{eq: prob_main_uncorr}.

\subsection{Equivalent optimization problem}

 We  express \eqref{eq: prob_main_uncorr}  in its epigraph form \cite[Sections\,3.1\&7.5]{boyvan04}
\begin{subequations}
\label{eq: prob_uncorr_epi}
\begin{empheq}{align}
 & \minimize   \quad   \mathbf 1^T \mathbf u    \label{eq: obj_uncorr} \\
& \st  \quad   \displaystyle   \frac{  \sigma_{\epsilon}^2    \mathbf w_k^T  \mathbf R_k \mathbf w_k  + \sigma_{\varsigma}^2 }{\mathbf w_k^T \mathbf S_k \mathbf w_k    +    \sigma_{\varsigma}^2 } \leq u_k, ~ k \in [K]  \label{eq: cons1_uncorr}\\
&\hspace*{0.73in}  \mathbf w^T \mathbf Q_{m}  \mathbf w  \leq E_{m},  ~ \hspace*{0.4in} m \in [M],  \label{eq: cons2_uncorr}
\end{empheq}
\end{subequations}
where $\mathbf u = [u_1, u_2, \ldots, u_K]^T$ is the vector of 
newly introduced optimization variables. 

We further introduce new variables $r_k$ and $s_k$ for $k \in [K]$ to rewrite \eqref{eq: cons1_uncorr} as
\begin{align}
\left \{
\begin{array}{l}
 \displaystyle  \frac{r_k}{s_k} \leq u_k, ~ s_k > 0  \vspace*{0.03in}  \\
 \mathbf w_k^T \mathbf S_k \mathbf w_k   +   \sigma_{\varsigma}^2 \geq s_k
  \vspace*{0.03in} \\
  \sigma_{\epsilon}^2    \mathbf w_k^T  \mathbf R_k \mathbf w_k   +  \sigma_{\varsigma}^2 \leq r_k ,
\end{array}
\right.
\label{eq: cons1_uncorr_v2}
\end{align}
where the equivalence between \eqref{eq: cons1_uncorr}  and \eqref{eq: cons1_uncorr_v2} holds since the minimization of $\mathbf 1^T \mathbf u$   with the above inequalities  forces the variable  $  s_k$ and  $r_k$  to   achieve their upper and lower  bounds, respectively. 

In \eqref{eq: cons1_uncorr_v2}, the ratio $r_k/s_k \leq u_k$ together with $s_k > 0$   can be reformulated as a quadratic inequality of DC type
\begin{align}
 s_k^2 + u_k^2 + 2 r_k -  (s_k + u_k)^2 \leq 0,
\label{eq: bilinear_quadratic}
\end{align}
where both $ s_k^2 + u_k^2 + 2 r_k $ and $(s_k + u_k)^2$ are convex quadratic functions.

From \eqref{eq: cons1_uncorr_v2} and \eqref{eq: bilinear_quadratic}, problem \eqref{eq: prob_uncorr_epi} becomes
\begin{subequations}
\label{eq: prob_uncorr_qp}
\begin{empheq}{align}
 & \minimize  \quad    \mathbf 1^T \mathbf u    \label{eq: obj_uncorr} \\
& \st    \quad   s_k^2 + u_k^2 + 2 r_k   \leq  (s_k + u_k)^2,~ k \in [K] \label{eq: cons1_qp}\\
&\hspace*{0.73in}     s_k -  \mathbf w_k^T \mathbf S_k \mathbf w_k   -   \sigma_{\varsigma}^2 \leq 0, ~ \hspace*{0.2in} k \in [K] \label{eq: cons2_qp} \\
&\hspace*{0.73in}     \sigma_{\epsilon}^2    \mathbf w_k^T  \mathbf R_k \mathbf w_k   +  \sigma_{\varsigma}^2 \leq r_k, ~ \hspace*{0.257in} k \in [K] \label{eq: cons3_qp} \\
&\hspace*{0.73in}  \mathbf w^T \mathbf Q_{m}  \mathbf w  \leq E_{m},  ~\hspace*{0.6in} m \in [M]  \label{eq: cons4_qp} \\
&  \hspace*{0.75in}  \mathbf s > \mathbf 0,
\end{empheq}
\end{subequations}
where the optimization variables are $\mathbf w$, $\mathbf u$, $\mathbf r$ and $\mathbf s$, $\mathbf r = [r_1, r_2, \ldots, r_K]^T$, $\mathbf s = [s_1, s_2, \ldots, s_K]^T$, and $>$   denotes elementwise  inequality.
Note that the quadratic functions of DC type in
\eqref{eq: cons1_qp} and \eqref{eq: cons2_qp} contain the nonconvexity of problem \eqref{eq: prob_uncorr_qp}.
In what follows, we will show that CCP is a suitable convex restriction approach for solving this problem.

\subsection{Convex restriction}

Problem \eqref{eq: prob_uncorr_qp} is convex except for the  nonconvex quadratic constraints \eqref{eq: cons1_qp} and \eqref{eq: cons2_qp}, which have the DC form
\begin{align}
f(\mathbf v)    - g(\mathbf v) \leq 0, \label{eq: DC_func}
\end{align}
where both $f $ and $g$ are convex functions. In \eqref{eq: cons1_qp}, we have 
$f(s_k, u_k, r_k) =   s_k^2 + u_k^2 + 2 r_k $, and $g(s_k, u_k) = (s_k + u_k)^2$. In \eqref{eq: cons2_qp},  $f( s_k) =   s_k$, and $g(\mathbf w_k) =  \mathbf w_k^T \mathbf S_k \mathbf w_k   +   \sigma_{\varsigma}^2 $.

We can convexify \eqref{eq: DC_func} by linearizing $g$ around a feasible point $\hat {\mathbf v}$, 
\begin{align}
f(\mathbf v) - \hat g(\mathbf v) \leq 0, \label{eq: DC_func_lin}
\end{align}
where $ \hat g(\mathbf v) \Def g(\hat {\mathbf v}) + ( \frac{\partial g(\hat {\mathbf v})}{\partial \mathbf v})^T (\mathbf v - \hat {\mathbf v})$, $\frac{\partial g(\hat {\mathbf v})}{\partial \mathbf v}$  is the
first-order derivative of $g$ at the point $\hat {\mathbf v}$. 
In \eqref{eq: DC_func_lin},  $\hat g$ is an affine \textit{lower bound} on the convex function $g$, and therefore, the set of $\mathbf v$ that satisfy \eqref{eq: DC_func_lin} is a strict \textit{subset} of the set of $\mathbf v$ that satisfy \eqref{eq: DC_func}. This implies that  a solution of the
 optimization problem with the linearized constraint \eqref{eq: DC_func_lin} is 
locally optimal for the problem with the original nonconvex
constraint \eqref{eq: DC_func}. 
 
We   can obtain a restricted convex version of problem
\eqref{eq: prob_uncorr_qp} by linearizing  
\eqref{eq: cons1_qp} and \eqref{eq: cons2_qp} as was done in \eqref{eq: DC_func} and   \eqref{eq: DC_func_lin}. We then solve a sequence of convex programs with iteratively updated linearization points. The use of
linearization to convexify nonconvex problems with DC type
functions  is known as CCP \cite{lipboy14}. At each iteration of CCP, we solve  
\begin{align}
\label{eq: prob_ucorr_lin}
\begin{array}{clr}
\displaystyle \minimize  &     \mathbf 1^T \mathbf u & \vspace*{0.03in}\\
\st &     s_k^2 + u_k^2 + 2 r_k - \hat g_1(s_k, u_k) \leq 0, &~ k \in [K] \vspace*{0.03in}\\
&  s_k - \hat g_2 (\mathbf w_k) \leq 0, &~ k \in [K]\vspace*{0.03in}\\
& \sigma_{\epsilon}^2    \mathbf w_k^T  \mathbf R_k \mathbf w_k   +  \sigma_{\varsigma}^2 \leq r_k,& ~ k \in [K]  \vspace*{0.03in}\\
&  \mathbf w^T \mathbf Q_{m}  \mathbf w  \leq E_{m},  &~m \in [M] \\
& \mathbf s > \mathbf 0,&
\end{array}
\hspace*{-0.2in}
\end{align}
where the optimization variables are $\mathbf w$, $\mathbf u$, $\mathbf r$, and $\mathbf s$,  $\hat g_1$ and $\hat g_2$ are  affine approximations  of $(s_k+u_k)^2$ and $\mathbf w_k^T \mathbf S_k \mathbf w_k   +   \sigma_{\varsigma}^2 $, namely,
$\hat g_1(s_k, u_k) \Def 2(\hat s_k + \hat u_k)(s_k + u_k) -  (\hat s_k + \hat u_k)^2$, and $\hat g_2(\mathbf w_k) \Def  2 \hat {\mathbf w}_k^T \mathbf S_k  {\mathbf w}_k -  \hat {\mathbf w}_k^T \mathbf S_k \hat {\mathbf w}_k  +   \sigma_{\varsigma}^2 $.  
We summarize CCP for solving problem \eqref{eq: prob_uncorr_qp} or \eqref{eq: prob_main_uncorr} in Algorithm\,1.

\begin{algorithm}
\caption{CCP for solving \eqref{eq: prob_main_uncorr}}
\begin{algorithmic}[1]
\Require initial points     $\hat {\mathbf w}$, $\hat { \mathbf s}$ and $\hat {\mathbf u}$, and  $\epsilon_{\mathrm{ccp}} > 0$  
\For{iteration $t =1,2,\ldots$}
\State solve problem \eqref{eq: prob_ucorr_lin} for the solution $(\mathbf w^{t}, \mathbf s^{t}, \mathbf u^t)$ 
\State update the linearization point, $\hat {\mathbf w} = \mathbf w^{t}$, 
$\hat { \mathbf s}= \mathbf s^{t}$, and 
\hspace*{0.22in}$\hat { \mathbf u}= \mathbf u^{t}$ 
\State \textbf{until} $| \mathbf 1^T \mathbf u^t - \mathbf 1^T \mathbf u^{t-1}| \leq \epsilon_{\mathrm{ccp}}$ with $t \geq 2$.
\EndFor
\end{algorithmic}
\end{algorithm}

To initialize Algorithm\,1,  we can choose  random   points, for example drawn
from a standard uniform distribution, that are then scaled to satisfy the constraints
   \eqref{eq: cons1_qp}\,--\,\eqref{eq: cons4_qp}.  Our extensive numerical examples show that Algorithm\,1  is  fairly  robust with respect to the choice of   the initial point; see Fig.\,\ref{fig: converge}-(a) for an example. 

{It is known from  \cite[Theorem\,10]{Sriperumbudur2009OnTC}  that CCP is a descent algorithm that converges to a stationary point of the original nonconvex problem. To be specific, at each iteration, we solve a  restricted
convex problem with a smaller feasible set which contains the
linearization point (i.e., the solution after the previous iteration). Therefore, we always obtain a new feasible point with a lower or equal objective value.
Moreover, reference \cite{ianpen12}  showed  that  CCP has \textit{at least}   linear convergence rate $O(1/t)$, where $t$ is the number of iterations\footnote{Given the stopping tolerance $\epsilon_{\mathrm{ccp}}$, the linear convergence rate implies $O(1/\epsilon_{\mathrm{ccp}})$ iterations to convergence.}.  
However, our numerical results and those in \cite{kuayuuts16,lipboy14,yuiran03} have shown that
the empirical convergence rate is typically faster, and much of the benefit of using  
CCP is gained during its first few iterations. } 

The computation cost of Algorithm\,1 is dominated by
the solution of the convex program with quadratic constraints
at Step\,2. This has the  computational complexity $O(a^3 + a^2 b)$ in the use of interior-point algorithm \cite[Chapter\,10]{nem12}, where $a$ and $b$ denote  the number of optimization variables and constraints, respectively. 
In problem \eqref{eq: prob_ucorr_lin}, we have $a =3K + KL $ and $b = 4K + M$. Therefore, the complexity of our algorithm is roughly given by $O( L^3)$ per iteration. Here we focus on the scenario in which the number of   collaboration links $L$ is much larger than $K$ or $M$.

\section{General Case: Optimal Sensor Collaboration for The  Estimation of  Correlated   Parameters}
\label{sec: LMMSE_gen}

 Different from    \eqref{eq: prob_main_uncorr},  
the presence of temporal correlation  
 makes finding the solution of \eqref{eq: prob_main_LMMSE}  more challenging. {However, we demonstrate that \eqref{eq: prob_main_LMMSE} can be recast as an optimization problem with the important property that the problem becomes a  semidefinite program (SDP) if its rank-one constraint is replaced by a linear relaxation/approximation.} Spurred by the problem structure, we employ a \textit{penalty  CCP} to solve \eqref{eq: prob_main_LMMSE}, and propose a fast optimization algorithm by using the alternating direction method of multipliers (ADMM).

\subsection{Equivalent optimization problem}

We transform \eqref{eq: prob_main_LMMSE} into the following equivalent form
\begin{align}
\begin{array}{llr}
\displaystyle \minimize 
& \tr (\mathbf V)   & \vspace*{0.03in}\\
\st & \mathbf P_w^{-1}   \succeq \mathbf V^{-1} & \vspace*{0.03in} \\
& \mathbf w^T \mathbf Q_{m} \mathbf w \leq E_{m},  &~ m \in [M],
\end{array}
\hspace*{-0.25in}\label{eq: prob_corr_v1}
\end{align}
where $\mathbf V \in \mathbb S^K$ 
is the newly introduced  optimization variable,
$\mathbb S^n$ represents  the  set of $n \times n $ symmetric  matrices,  and  the  notation $\mathbf  X \succeq \mathbf Y$ (or $\mathbf X \preceq \mathbf Y$)
indicates that $\mathbf X - \mathbf Y$ (or $\mathbf Y - \mathbf X$) is positive semidefinite.  The first inequality constraint of problem \eqref{eq: prob_corr_v1} is obtained from $\mathbf P_w  \preceq \mathbf V$, where $\mathbf P_w$ is given by
\eqref{eq: MSE_LMMSE_final}, and $ \mathbf P_w^{-1}  $ represents the Bayesian Fisher information  matrix. 

 We further
  introduce  a new vector of optimization variables   $\mathbf p = [p_1, \ldots, p_K]^T$   such that the first matrix inequality of problem \eqref{eq: prob_corr_v1} is expressed as
\begin{align}
&\mathbf C - \diag(\mathbf p)
 \succeq \mathbf V^{-1},  \label{eq: LMI1} \\
&p_k \geq \sigma_{\epsilon}^{-2}  \mathbf h_k^T \left ( \mathbf I +\sigma_{\epsilon}^{2}  \sigma_{\varsigma}^{-2}  \mathbf G_k^T \mathbf U_k \mathbf G_k \right )^{-1}  \mathbf h_k,  ~k \in [K],  \label{eq: LMI2} \\
& \mathbf U_k = \mathbf w_k \mathbf w_k^T, \label{eq: Uww}
\end{align}
where
we use the expression of $\mathbf P_w$ given by \eqref{eq: MSE_LMMSE_final}, and 
$\mathbf U_k \in \mathbb S^L$ is the newly introduced  optimization variable for $k \in [K] $.
Note that the minimization of
$\tr(\mathbf V)$ with inequalities \eqref{eq: LMI1} and
\eqref{eq: LMI2}  would  force  the  variable  $p_k$ to  achieve  its  lower  bound. In  other  words,  problem  \eqref{eq: prob_corr_v1} is  equivalent  to  the  problem in  which  the  first inequality  constraint  of  \eqref{eq: prob_corr_v1}  is  replaced  by  the above
two  inequalities. 

By employing  the  Schur
complement, we can express \eqref{eq: LMI1} and
\eqref{eq: LMI2} as
the   linear matrix inequalities (LMIs)
\begin{align}
&\begin{bmatrix}
\mathbf C -\diag (\mathbf p)   & \mathbf I \\
\mathbf I &  \mathbf V
\end{bmatrix} \succeq 0, \label{eq: LMIs1} \\
&\begin{bmatrix} 
p_k & \sigma_{\epsilon}^{-1}  \mathbf h_k^T\\
\sigma_{\epsilon}^{-1}  \mathbf h_k &  \mathbf I +\sigma_{\epsilon}^{2}  \sigma_{\varsigma}^{-2}  \mathbf G_k^T \mathbf U_k \mathbf G_k 
\end{bmatrix} \succeq 0,~ k \in [K].
\label{eq: LMIs2}
\end{align}

Replacing the first   inequality of problem \eqref{eq: prob_corr_v1} with LMIs
 \eqref{eq: LMIs1}\,--\,\eqref{eq: LMIs2},  
we obtain an optimization problem that is convex except for the rank-one constraint \eqref{eq: Uww}, which
can be recast as two inequalities
 \begin{align}
  \mathbf U_k - \mathbf w_k\mathbf w_k^T  \succeq 0, ~
 \mathbf U_k - \mathbf w_k \mathbf w_k^T \preceq 0, ~ k \in [K].
\label{eq: ineqs_rank1}
\end{align}
According to the Shur complement, the first matrix inequality is equivalent to the LMI
\begin{align}
\begin{bmatrix}
\mathbf U_k & \mathbf w_k  \\
\mathbf w_k^T & 1
\end{bmatrix} \succeq 0, ~ k \in [K].
\label{eq: LMI1_nx}
\end{align}
And the second  inequality in   \eqref{eq: ineqs_rank1} involves a function of DC type, where  $\mathbf U_k$ and $\mathbf w_k \mathbf w_k^T$ are matrix convex   functions  \cite{boyvan04}. 

From \eqref{eq: LMIs1}\,--\,\eqref{eq: LMI1_nx},  problem \eqref{eq: prob_corr_v1} or \eqref{eq: prob_main_LMMSE} is  equivalent to
\begin{subequations}
\label{eq: prob_corr_v2}
\begin{empheq}{align}
&\displaystyle \minimize 
  \quad \tr (\mathbf V)    \\
&\st  \quad  \mathbf w^T \mathbf Q_{m} \mathbf w \leq E_{m},  \quad m \in [M] \label{cons: ene_LMMSE}\\ 
& \hspace*{0.73in}
\text{LMIs in (\ref{eq: LMIs1})\,--\,(\ref{eq: LMIs2})} \\ 
& \hspace*{0.73in} \text{LMIs in (\ref{eq: LMI1_nx})}  \label{cons: LMI_nx} \\
& \hspace*{0.73in} \mathbf U_k - \mathbf w_k \mathbf w_k^T \preceq 0, \quad   k \in [K], \label{cons: DC_nx}
\end{empheq}
\end{subequations}
where the optimization variables are $\mathbf w$, $\mathbf p$, $\mathbf V$ and $\mathbf U_k$ for $k \in [K]$, and \eqref{cons: DC_nx} is  a nonconvex constraint of DC type.

\subsection{Convexification}

Proceeding  with  the  same  logic  as  in  Sec.\,\ref{sec: col_uncorr} to convexify the constraint \eqref{eq: DC_func},  we linearize \eqref{cons: DC_nx} around a point $\hat {\mathbf w}_k$,
\begin{align}
\mathbf U_k - \hat {\mathbf w}_k \mathbf w_k^T -   \mathbf w_k \hat {\mathbf w}_k^T + \hat {\mathbf w}_k \hat {\mathbf w}_k^T \preceq 0, ~ k \in [K].
\label{eq: DC_lin}
\end{align}
 It is  straightforward to apply CCP to solve problem \eqref{eq: prob_corr_v2}  by replacing  \eqref{cons: DC_nx} with \eqref{eq: DC_lin}. However, such an approach 
fails in practice. This is not surprising, since the feasible set determined by \eqref{cons: LMI_nx} and \eqref{eq: DC_lin}    only contains the linearization point. 
Specifically, from \eqref{cons: LMI_nx} and \eqref{eq: DC_lin}, we obtain  
\begin{align}
& (\mathbf w_k - \hat {\mathbf w}_k) (\mathbf w_k - \hat {\mathbf w}_k)^T \nonumber \\
 = &~ \mathbf w_k \mathbf w_k^T - \hat {\mathbf w}_k \mathbf w_k^T -   \mathbf w_k \hat {\mathbf w}_k^T + \hat {\mathbf w}_k \hat {\mathbf w}_k^T  \nonumber \\
 \preceq & ~\mathbf U_k - \hat {\mathbf w}_k \mathbf w_k^T -   \mathbf w_k \hat {\mathbf w}_k^T + \hat {\mathbf w}_k \hat {\mathbf w}_k^T ~ \preceq ~ 0,
\end{align}
which indicates that $\mathbf w_k  = \hat {\mathbf w}_k$. Therefore, CCP 
  gets trapped in the linearization point. 

 \begin{remark}{
Dropping the nonconvex constraint \eqref{cons: DC_nx} is another  method  to convexify  problem \eqref{eq: prob_corr_v2}, known as semidefinite relaxation \cite{luomasoyezha10}. 
However, such an approach makes 
  the optimization variable  $\mathbf U_k$ unbounded, since the minimization of $\tr(\mathbf V)$ forces $\mathbf U_k$ to be as large as possible such that 
the variable $p_k$ in \eqref{eq: LMI2} is as small as possible.}
\end{remark}

In order to circumvent the drawback of  the standard CCP, we consider its penalized version,  known as penalty CCP    \cite{lipboy14,farjov14_ACC}, where we add   new variables to allow for   constraints   \eqref{eq: DC_lin} to be violated and penalize the sum of the violations in the objective function. As a result, the convexification   \eqref{eq: DC_lin}  is modified by
\begin{align}
\mathbf U_k - \hat {\mathbf w}_k \mathbf w_k^T -   \mathbf w_k \hat {\mathbf w}_k^T + \hat {\mathbf w}_k \hat {\mathbf w}_k^T \preceq \mathbf Z_k, ~k \in [K],
\label{eq: DC_lin_penalty}
\end{align}
where $\mathbf Z_k \in \mathbb S^L$ is a newly introduced variable. 
The constraint \eqref{eq: DC_lin_penalty} implicitly adds the additional constraint $\mathbf Z_k \succeq 0$ due to $\mathbf U_k \succeq \mathbf w_k \mathbf w_k$  from \eqref{cons: LMI_nx}. 

After replacing \eqref{cons: DC_nx} with \eqref{eq: DC_lin_penalty}, we obtain the SDP,
\begin{align}
\begin{array}{ll}
 \minimize
 & \displaystyle \tr (\mathbf V)  + \tau \sum_{k=1}^K \tr(\mathbf Z_k) \vspace*{0.03in} \\
\st 
&  \text{\eqref{cons: ene_LMMSE}\,--\,\eqref{cons: LMI_nx}  and \eqref{eq: DC_lin_penalty}} 
\end{array}
\label{eq: prob_corr_penalty}
\end{align}
where the optimization variables are $\mathbf w$, $\mathbf p$, $\mathbf V$, $\mathbf U_k$ and $\mathbf Z_k $ for $k \in [K]$, and $\tau > 0$
is a penalty parameter. Compared to the standard CCP, problem \eqref{eq: prob_corr_penalty} is optimized over a larger feasible set
  since we allow for constraints to be violated by adding   variables $\mathbf Z_k$ for $k \in [K]$. We summarize the use of penalty CCP to solve \eqref{eq: prob_main_LMMSE} in Algorithm\,2.

\begin{algorithm}
\caption{Penalty CCP for  solving \eqref{eq: prob_main_LMMSE}}
\begin{algorithmic}[1]
\Require 
 an initial point  $\hat {\mathbf w}$,  $\epsilon_{\mathrm{ccp}} > 0$, $\tau^{0} > 0$, $\tau_{\max}>0$  and $\mu > 1$.
\For{iteration $t=1,2,\ldots$}
\State solve problem \eqref{eq: prob_corr_penalty} for its solution $\mathbf w^{t}$ via SDP solver 
\hspace*{0.21in}or   ADMM-based algorithm in Sec.\,\ref{sec: ADMM}
\State update the linearization point, $\hat {\mathbf w} = \mathbf w^{t}$
\State update the penalty parameter $\tau^t = \min \{ \mu \tau^{t-1},\tau_{\max}\}$
\State let $\psi^t$ be the objective value of \eqref{eq: prob_corr_penalty}
\State \textbf{until} $| \psi^{t} - \psi^{t-1} | \leq \epsilon_{\mathrm{ccp}}$ with $t \geq 2$.
\EndFor
\end{algorithmic}
\end{algorithm}

In Algorithm\,2, the initial point $\hat {\mathbf w}$ is randomly picked from   a standard uniform distribution. Note that   $\hat {\mathbf w}$ is not necessarily    feasible for  \eqref{eq: prob_main_LMMSE} since  violations of   constraints are allowed. {We also remark that once $\tau = \tau_{\max}$ (after at most $\log_{\mu}(\tau_{\max}/\tau_0)$ iterations), the penalty CCP reduces to CCP. Therefore, the penalty CCP enjoys the same convergence properties of CCP.} 

The  computation cost of  Algorithm\,2  is  dominated  by the solution of the SDP  \eqref{eq: prob_corr_penalty}  at Step\,2.  
This leads to the complexity
$O(a^2 b^{2} + a b^{3})$  by using the interior-point alogrithm in   off-the-shelf  solvers \cite[Chapter\,11]{nem12}, where
$a$ and $b$  are the number of optimization variables and the size of the semidefinite matrix, respectively. 
In  \eqref{eq: prob_corr_penalty},  the number of optimization variables is proportional to $L^2$. Therefore, 
the complexity of Algorithm\,2 is roughly given by  $O(L^{6})$.
Clearly,  computing solutions to SDPs  becomes inefficient for problems  of medium or large size. 
In what follows, we will develop an ADMM-based algorithm that is   more amenable to large-scale optimization.



\subsection{Fast algorithm via ADMM}
\label{sec: ADMM}
It has been shown in \cite{boyparchupeleck11,parboy13,odochuparboy13,shizhaodo15} that ADMM is a powerful tool for solving large-scale optimization problems. {The major advantage of ADMM is that it allows us to split the original   problem  into subproblems, each of which   can be solved more   efficiently or even analytically. In what follows, we will employ ADMM to solve problem \eqref{eq: prob_corr_penalty}.}

It is shown in Appendix\,\ref{appendix: ADMM} that problem  \eqref{eq: prob_corr_penalty}  can be reformulated in a way  that lends itself to the application of ADMM.  This is achieved by introducing     slack variables  and indicator functions    to express the inequality constraints of problem    \eqref{eq: prob_corr_penalty} as linear equality constraints together with     cone constraints with respect to slack variables, including second-order cone     and positive semidefinite cone constraints.

ADMM is   performed based on the augmented Lagrangian  \cite{boyparchupeleck11}  of the reformualted problem \eqref{eq: prob_corr_penalty}, and leads to   two problems,   the first of which can be treated as an unconstrained quadratic program and the latter renders an analytical solution.   
These two problems are solved iteratively
and ‘communicate’ to each other through special quadratic
terms in their objectives; the quadratic term in each problem
contains information about the solution of the other problem
and also about   dual variables (also known as Lagrange
multipliers). In what follows, we refer to these
problems as the `$\mathscr X$-minimization' and `$\mathscr Z$-minimization' problems.
Here  $ \mathscr X $ denotes  the  set   of  {primal  variables}  
$\mathbf w$, $\mathbf p$, $\mathbf V$, $\mathbf U_k$ and $\mathbf Z_k $ for $k \in [K]$, and $ \mathscr Z$ denotes   the  set  of  {slack   variables}  
$\boldsymbol \lambda_m$, $\boldsymbol \Lambda_1 $ and $ \{ \boldsymbol \Lambda_{i,k} \}_{i=2,3,4} $   for $m \in [M]$ and $k \in [K]$. We also use $\mathscr Y$ to denote the set of dual variables   $\boldsymbol \pi_m$, $\boldsymbol \Pi_1$ and $\{ \boldsymbol \Pi_{i,k} \}_{i=2,3,4}$ for $m \in [M]$  and $k \in [K]$.    
The   ADMM algorithm is precisely described by  \eqref{eq: p_step}\,--\,\eqref{eq: admm_3steps} in Appendix\,\ref{appendix: ADMM}.  
 
We emphasize that the crucial property of the  ADMM approach is that, as
we demonstrate in the rest of this section, the solution of
each of the $\mathscr X$- and $\mathscr Z$-minimization problems can be found
exactly and efficiently.

\subsubsection{$\mathscr X$-minimization step} 
The $\mathscr X$-minimization problem can be cast as
\begin{align}
\minimize & ~~ \varphi(\mathbf w, \mathbf p, \mathbf V, \{ \mathbf U_k \}, \{ \mathbf Z_k \}). \label{eq: UQP}
\end{align}
The objective function of problem \eqref{eq: UQP} is given by
 \eqref{eq: UQP_obj}, where
$\boldsymbol \alpha_m \Def \boldsymbol \lambda_m^t - \mathbf c_m - (1/\rho) \boldsymbol \pi_m^t$ for $m \in [M]$, $\boldsymbol \Upsilon_1 \Def \boldsymbol \Lambda_1^t - (1/\rho) \boldsymbol \Pi_1^t$, and $\boldsymbol \Upsilon_{i,k} \Def \boldsymbol \Lambda_{i,k}^t - (1/\rho) \boldsymbol \Pi_{i,k}^t $ for $i \in \{ 2,3,4\}$ and $k \in [K]$, and $t$ denotes the ADMM iteration. For ease of notation, 
we will omit the ADMM iteration index $t$ in what follows.

\begin{figure*}[!htb]
\begin{align}
\label{eq: UQP_obj}
  \varphi(\mathbf w, \mathbf p, \mathbf V, \{ \mathbf U_k \}, \{ \mathbf Z_k \}) \Def & \tr (\mathbf V)  + \tau \sum_{k=1}^K \tr(\mathbf Z_k) + \frac{\rho}{2} \sum_{m = 1}^M \left \| \bar {\mathbf Q}_m \mathbf w - \boldsymbol \alpha_m  \right \|_2^2 +  \frac{\rho}{2} \left \|
\begin{bmatrix}
\mathbf C -\diag (\mathbf p)   & \mathbf I \\
\mathbf I &  \mathbf V
\end{bmatrix} -\boldsymbol \Upsilon_1
 \right \|_F^2 \nonumber \\
&+ \frac{\rho}{2}  \sum_{k=1}^K  \left \| 
\begin{bmatrix} 
p_k & \sigma_{\epsilon}^{-1}  \mathbf h_k^T\\
\sigma_{\epsilon}^{-1}  \mathbf h_k &  \mathbf I +\sigma_{\epsilon}^{2}  \sigma_{\varsigma}^{-2}  \mathbf G_k^T \mathbf U_k \mathbf G_k 
\end{bmatrix} - \boldsymbol \Upsilon_{2,k} 
\right \|_F^2 + \frac{\rho}{2} \sum_{k=1}^K   \left \| 
\begin{bmatrix}
\mathbf U_k & \mathbf w_k  \\
\mathbf w_k^T & 1
\end{bmatrix}  - \boldsymbol \Upsilon_{3,k} 
\right \|_F^2 \nonumber \\
&+  \frac{\rho}{2}  \sum_{k=1}^K  \left \| 
\mathbf Z_k -  \mathbf U_k + \hat {\mathbf w}_k \mathbf w_k^T + \mathbf w_k \hat {\mathbf w}_k^T - \hat {\mathbf w}_k \hat {\mathbf w}_k^T - \boldsymbol \Upsilon_{4,k} 
\right \|_F^2
\end{align}
\hrulefill
\end{figure*}

We note that problem \eqref{eq: UQP} is an  unconstrained quadratic program (UQP) with   large amounts of variables. 
In order to reduce the computational complexity and memory requirement in optimization, 
we  will employ a gradient descent method \cite{boyvan04} together with a backtracking line search \cite[Chapter\,9.2]{boyvan04}  to solve this UQP.
In Proposition\,\ref{prop: gradient},
we show  the gradient of the objective function  of problem  \eqref{eq: UQP}. 

\begin{myprop}
\label{prop: gradient}
The gradient of the objective function of  problem \eqref{eq: UQP} is given by
\begin{align}
\begin{array}{l}
  \nabla_{\mathbf w} \varphi =    \rho \sum_{m=1}^M \bar {\mathbf Q}_m^T  
(\bar {\mathbf Q}_m \mathbf w - \boldsymbol \alpha_m ) +  2 \rho (\mathbf w - \boldsymbol \gamma_3)  \vspace*{0.03in}  \\
  \hspace*{0.45in}+ 2 \rho \,  \mathrm{blkdiag}\{\hat {\mathbf w}_k \mathbf w_k^T + \mathbf w_k \hat {\mathbf w}_k^T - \mathbf H_k \}_{k=1}^K \hat{\mathbf w}   \vspace*{0.05in}\\
 \nabla_{\mathbf p} \varphi =        2 \rho  \mathbf p + \rho ( \diag(\boldsymbol \Upsilon_1^{11} )  -\diag(\mathbf C)  - \boldsymbol \gamma_2)   \vspace*{0.05in} \\
  \nabla_{\mathbf V} \varphi =  \mathbf I + \rho(\mathbf V - \boldsymbol \Upsilon_1^{22})   \vspace*{0.05in} \\
 \nabla_{\mathbf U_k} \varphi =   { \rho\sigma_{\epsilon}^{2} }{\sigma_{\varsigma}^{-2}  } \mathbf G_k
(\mathbf I + {  \sigma_{\epsilon}^{2} }{\sigma_{\varsigma}^{-2}  } \mathbf G_k^T \mathbf U_k \mathbf G_k  - \boldsymbol \Upsilon_{2,k}^{22})   \mathbf G_k^T    \\
 \hspace*{0.5in} + \rho (2 \mathbf U_k - \boldsymbol \Upsilon_{3,k}^{11} -\mathbf Z_k-  \mathbf T_k),~ k \in [K]\vspace*{0.05in}   \\
  \nabla_{\mathbf Z_k} \varphi =  \tau \mathbf I + \rho (\mathbf Z_k - \mathbf U_k + \mathbf T_k), ~ k \in [K],
\end{array}
\nonumber
\end{align}
where $\boldsymbol \gamma_3 = [ \boldsymbol \gamma_{3,1}^T, \ldots, \boldsymbol \gamma_{3,K}^T]^T$, $\boldsymbol \gamma_{3,k} $ is the $(L+1)$ column of  $\boldsymbol \Upsilon_{3,k}$ after the last entry is removed,  $\mathbf H_k \Def \mathbf U_k - \mathbf Z_k +  \hat {\mathbf w}_k \hat {\mathbf w}_k^T + \boldsymbol \Upsilon_{4,k} $,
$\hat{\mathbf w} = [\hat{\mathbf w}_1^T, \ldots, \hat{\mathbf w}_K^T]^T$, 
$\boldsymbol \Upsilon_1^{11}$ is a submatrix  of  $\boldsymbol \Upsilon_1$ that contains its first $K$  rows and columns,    $\boldsymbol \gamma_2 = [\gamma_{2,1} , \ldots, \gamma_{2,K}]^T$, $\gamma_{2,k}$ is    the first element of $\boldsymbol \Upsilon_{2,k}$,  $\diag(\cdot)$ returns  the diagonal entries of its matrix argument in vector form, $\boldsymbol \Upsilon_1^{22}$ is a submatrix of  $\boldsymbol \Upsilon_1$ after the first $K$ rows and columns are removed,  $\boldsymbol \Upsilon_{2,k}^{22} $ is   a submatrix of  $\boldsymbol \Upsilon_{2,k}$ after the first  row and column  are removed, $\boldsymbol \Upsilon_{3,k}^{11}$ is a   submatrix of  $\boldsymbol \Upsilon_{3,k}$ after the last  row and column  are removed,   and $\mathbf T_k \Def \hat {\mathbf w}_k \mathbf w_k^T + \mathbf w_k \hat {\mathbf w}_k^T - \hat {\mathbf w}_k \hat {\mathbf w}_k^T - \boldsymbol \Upsilon_{4,k} $.
\end{myprop}
\textbf{Proof:} See Appendix\,\ref{app: gradient}. \hfill $\blacksquare$

In Proposition\,\ref{prop: gradient}, the optimal values  of $\mathbf p$ and $\mathbf V$ are achieved by letting $\nabla_{\mathbf p} \varphi = \mathbf 0$ and $\nabla_{\mathbf V} \varphi = \mathbf 0$, which yield
\begin{align}
\mathbf p = \frac{1}{2}(  \diag(\mathbf C)  + \boldsymbol \gamma_2 - \diag(\boldsymbol \Upsilon_1^{11} )   ),~\mathbf V = \boldsymbol \Upsilon_1^{22} - \frac{1}{\rho} \mathbf I. 
\label{eq: V_opt}
\end{align}
To solve problem \eqref{eq: UQP} for other variables, we employ the gradient descent method  
summarized in Algorithm\,3. This algorithm calls on the backtracking line search (Algorithm\,4)  to properly determine   the step size such that  
the convergence to a stationary point of problem \eqref{eq: UQP} is accelerated.

\begin{algorithm}
\caption{Gradient descent method for solving UQP \eqref{eq: UQP}}
\begin{algorithmic}[1]
\Require  values of  $\mathbf w$,  $\{ \mathbf U_k \}$ and $\{ \mathbf Z_k \}$   at the previous ADMM iteration,    $\mathbf p$ and $\mathbf V$ given by \eqref{eq: V_opt}, and  $\epsilon_{\mathrm{grad}}  > 0$
\Repeat
\State compute the gradient of $\phi$ following  Proposition\,\ref{prop: gradient} 
\State compute $ c_{\mathrm {grad}} \Def \sum_{k=1}^K  \|\nabla_{\mathbf U_k} \varphi  \|_F^2+   \| \nabla_{\mathbf w} \varphi \|_2^2    $ 
\hspace*{0.2in}$  + \sum_{k=1}^K   \|\nabla_{\mathbf Z_k} \varphi \|_F^2 
$ 
\State call Algorithm\,4 to determine a step size $\kappa$ 
\State update variables 
$
\mathbf w \Def \mathbf w + \kappa    \nabla_{\mathbf w} \varphi,~
\mathbf U_k \Def \mathbf U_k +$ \hspace*{0.2in} $\kappa    \nabla_{\mathbf U_k} \varphi,~
\mathbf Z_k \Def \mathbf Z_k + \kappa    \nabla_{\mathbf Z_k} \varphi
$
\Until 
$c_{\mathrm {grad}}   \leq \epsilon_{\mathrm{grad}}. $
\end{algorithmic}
\end{algorithm}
\begin{algorithm}
\caption{Backtracking line search for choosing  $\kappa$}
\begin{algorithmic}[1]
\State Given $\kappa \Def 1$, $a_1 \in (0, 0.5$), $a_2 \in (0,1)$, and $c_{\mathrm {grad}} $
\Repeat
\State $\kappa \Def a_2 \kappa$,
\State let $\hat \varphi  $ be the value of $\varphi$ at the points 
$\mathbf w + \kappa    \nabla_{\mathbf w} \varphi $,
\hspace*{0.2in}$\mathbf U_k + \kappa    \nabla_{\mathbf U_k} \varphi 
$, and
$
\mathbf Z_k + \kappa    \nabla_{\mathbf Z_k} \varphi
$ 
\Until  
$\hat \varphi < \varphi(\mathbf w, \{ \mathbf U_k \}, \{ \mathbf Z_k \})   - a_1 \kappa \, c_{\mathrm {grad}}   $.
\end{algorithmic}
\end{algorithm}

\subsubsection{$\mathscr Z$-minimization step}
The $\mathscr Z$-minimization problem
is decomposed   with respect to  each of slack variables. 

$\bullet$\, {Subproblem with respect to $\boldsymbol \lambda_m$}:
\begin{align}
\begin{array}{ll}
\minimize & \| \boldsymbol \lambda_m - \boldsymbol \beta_m \|_2^2 \\
\st & \| [\boldsymbol \lambda_m]_{1:KL} \|_2 \leq [\boldsymbol \lambda_m]_{KL+1},
\end{array}
\label{eq: prob_socp_m}
\end{align}
where $\boldsymbol \beta_m \Def \bar {\mathbf Q}_m \mathbf w^{t+1} + \mathbf c_m + (1/\rho) \boldsymbol \pi_m^t$, and  $t$ is the ADMM iteration index. 
For   notational simplicity,    the ADMM iteration will be omitted in what follows.
The solution of problem 
\eqref{eq: prob_socp_m} is achieved by projecting $ \boldsymbol \beta_m $ onto a second-order cone  \cite[Sec.\,6.3]{parboy13},
\begin{align}
\boldsymbol \lambda_m = \left \{
\begin{array}{ll}
\mathbf 0 & \| [\boldsymbol \beta_m]_{1:KL}  \|_2 \leq    - [\boldsymbol \beta_m ]_{KL+1}  \\
\boldsymbol \beta_m  & \| [\boldsymbol \beta_m]_{1:KL}  \|_2 \leq    [\boldsymbol \beta_m]_{KL+1} \\
\tilde{\boldsymbol \beta}_m   & \|  [\boldsymbol \beta_m]_{1:KL}  \|_2 \geq    | [\boldsymbol \beta_m]_{KL+1}  |,
\end{array}
\right. 
\label{eq: sol_proj_socp}
\end{align}
for $m \in [M]$, where  
\[
\tilde{\boldsymbol \beta}_m \hspace*{-0.03in}  =  \hspace*{-0.03in} \frac{1}{2}\left (1 \hspace*{-0.03in}+ \hspace*{-0.03in} \frac{  [\boldsymbol \beta_m]_{KL+1}}{ \| [\boldsymbol \beta_m]_{1:KL} \|_2} \right ) \left [ [\boldsymbol \beta_m]_{1:KL}^T,  \| [\boldsymbol \beta_m]_{1:KL}   \|_2 \right ]^T.
\]

$\bullet$\, Subproblem with respect to $\boldsymbol \Lambda_1$:
\begin{align}
\begin{array}{ll}
\minimize & \|\boldsymbol \Lambda_1 - \boldsymbol \Phi_1 \|_F^2 \\
\st & \boldsymbol \Lambda_1  \succeq 0,
\end{array}
\label{eq: prob_SDP_1}
\end{align}
where $ \boldsymbol \Phi_1 \Def \begin{bmatrix}
\mathbf C -\diag (\mathbf p)   & \mathbf I \\
\mathbf I &  \mathbf V
\end{bmatrix} +(1/\rho)\boldsymbol \Pi_1$.
The solution of problem \eqref{eq: prob_SDP_1}  is given by  \cite[Sec.\,6.3]{parboy13}
\begin{align}
\boldsymbol \Lambda_1 =  
\sum_{i=1}^{2K} (\sigma_i)_+ \boldsymbol \omega_i  \boldsymbol \omega_i^T,
\label{eq: sol_proj_sdp}
\end{align}
where $\sum_{i=1}^{2K} \sigma_i \boldsymbol \omega_i  \boldsymbol \omega_i^T $ is the eigenvalue decomposition of $\boldsymbol \Phi_1$,  and $(\cdot)_+$
is the positive part operator. 

$\bullet$\, Subproblem with respect to $\boldsymbol \Lambda_{i,k}$ for $i \in \{ 2,3,4\}$ and $k \in [K]$:
\begin{align}
\begin{array}{ll}
\minimize & \|\boldsymbol \Lambda_{i,k} - \boldsymbol \Phi_{i,k} \|_F^2 \\
\st & \boldsymbol \Lambda_{i,k}  \succeq 0,
\end{array}
\label{eq: prob_SDP_i}
\end{align}
where
\begin{align}
\hspace*{-0.1in}
\begin{array}{l}
\boldsymbol \Phi_{2,k} \Def
\begin{bmatrix} 
p_k & \sigma_{\epsilon}^{-1}  \mathbf h_k^T\\
\sigma_{\epsilon}^{-1}  \mathbf h_k &  \mathbf I +\sigma_{\epsilon}^{2}  \sigma_{\varsigma}^{-2}  \mathbf G_k^T \mathbf U_k \mathbf G_k 
\end{bmatrix} + \frac{1}{\rho} \boldsymbol \Pi_{2,k} \\
\boldsymbol \Phi_{3,k} \Def
\begin{bmatrix}
\mathbf U_k & \mathbf w_k  \\
\mathbf w_k^T & 1
\end{bmatrix}  + \frac{1}{\rho} \boldsymbol \Pi_{3,k} \\
\boldsymbol \Phi_{4,k} \Def  \mathbf Z_k -  \mathbf U_k + \hat {\mathbf w}_k \mathbf w_k^T + \mathbf w_k \hat {\mathbf w}_k^T - \hat {\mathbf w}_k \hat {\mathbf w}_k^T + \frac{1}{\rho} \boldsymbol \Pi_{4,k}. \nonumber
\end{array}
\end{align}
The  solution of problem \eqref{eq: prob_SDP_i} is the same as \eqref{eq: sol_proj_sdp} except that $\boldsymbol \Phi_1$ is replaced with  $\boldsymbol \Phi_{i,k}$ for   $i \in \{ 2,3,4\}$ and $k \in [K]$.

\subsubsection{Summary of the proposed ADMM algorithm}

We initialize the ADMM algorithm by setting $\mathbf w^0 = \mathbf 1$, $\mathbf p^0 = \mathbf 1$, $\mathbf V^0 = \mathbf I$, $\mathbf U_k^0 = \mathbf Z_k^0 = \mathbf I$ for $k \in [K]$, $\boldsymbol \lambda_m^0 = \boldsymbol \pi_m^0 = \mathbf 0$ for $m \in [M]$, $\boldsymbol \Lambda_1^0 =\boldsymbol \Pi_1^0 =\mathbf 0 $, and $\boldsymbol \Lambda_{i,k}^0 =\boldsymbol \Pi_{i,k}^0 =\mathbf 0 $ for $i \in \{ 2,3,4,\}$ and $k \in [K]$. 
The ADMM approach is summarized in Algorithm\,5.
\begin{algorithm}
\caption{ADMM for solving problem \eqref{eq: prob_corr_penalty}}
\begin{algorithmic}[1]
\State Initialize variables and set $\rho$ and $\epsilon_{\mathrm{admm}}$
\For{iteration $t=1,2,\ldots$}
\State obtain optimal values of primal variables  $\mathscr X^t$  using
 \hspace*{0.2in}Algorithm\,3 and \eqref{eq: V_opt}
\State obtain  optimal values of slack variables  $\mathscr Z^t$   using \eqref{eq: sol_proj_socp}, \hspace*{0.2in}\eqref{eq: sol_proj_sdp} and \eqref{eq: prob_SDP_i}
\State update dual variables  based on \eqref{eq: admm_3steps}
\State \textbf{until} both $\| \mathscr X^{t+1} - \mathscr Z^{t} \|_F $  and 
$\|  \mathscr Z^{t+1} - \mathscr Z^{t } \|_F $ are 
 \hspace*{0.2in}less than  $ \epsilon_{\mathrm{admm}}$. 
\EndFor
\end{algorithmic}
\end{algorithm}

{The global convergence of ADMM has
been widely studied in \cite{heyuan12,denyin16,honluo13_TechRep}. It is   known from  \cite{heyuan12,denyin16,honluo13_TechRep} that   ADMM  has a linear convergence rate $O(1/t)$  for general convex optimization problems such as problem \eqref{eq: prob_corr_penalty}, where $t$ is the number of iterations. 
In practice, our numerical results and those in \cite{boyparchupeleck11,parboy13,odochuparboy13,shizhaodo15} have shown that ADMM can converge to
modest accuracy--sufficient for many applications--within a few tens of iterations. }


At each iteration of ADMM,  the computational  complexity of   the $\mathscr X$-minimization step  is approximated by
 $O(L^4)$, where $O(L )$ roughly counts for the number of iterations of the gradient descent method, and $O(L^3)$ is the complexity of matrix multiplication while computing  the gradient. Here we assume that $L $ is much larger than $K$ and $N$.  In $\mathscr Z$-minimization step, the computational complexity is dominated by the  eigenvalue decomposition used in  \eqref{eq: sol_proj_sdp}. This leads to the complexity $O(L^{3.5})$. As a result, the total computation cost of the   ADMM algorithm  is   given by $O(L^4)$.
{For  additional  perspective,  we  compare  the  computational
complexity  of the ADMM algorithm with  the interior-point algorithm that takes complexity $O(L^{6})$.
The complexity of ADMM decreases significantly in terms of the number of collaboration links by a factor $L^2$.  
We refer the reader to Sec.\,\ref{sec: numerical} for numerical results on the running time  improvement.
%
}

\section{Numerical Results}
\label{sec: numerical}
\begin{figure}[htb]
\centering
\includegraphics[width=.48\textwidth,height=!]{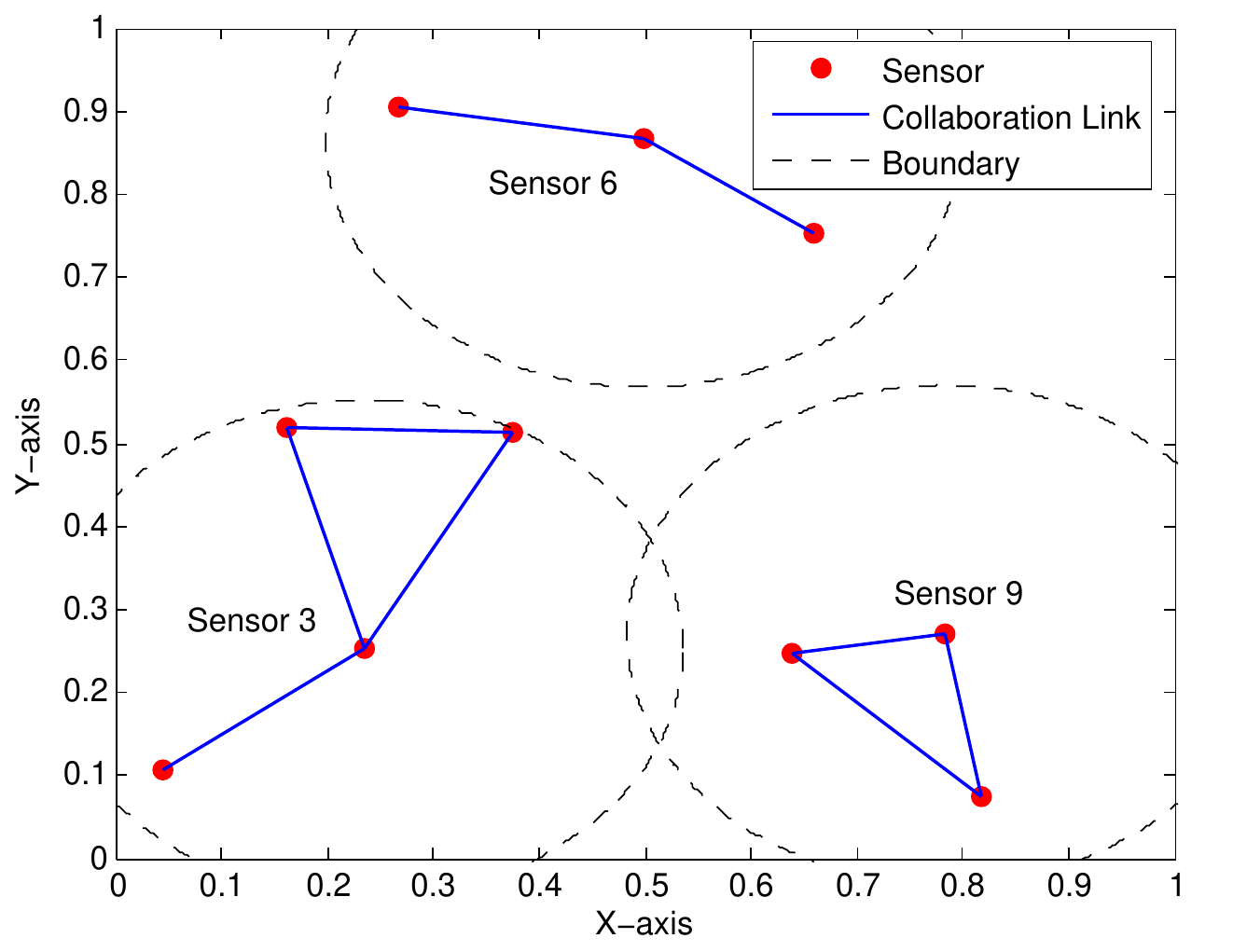} 
\caption{\footnotesize{$\mathrm{RGG}(10,0.3)$,  collaboration is depicted for sensors $3$, $6$ and $9$.
}}
  \label{fig: RGG}
\end{figure}

This section empirically shows   the effectiveness of our
approach for sensor collaboration in time-varying sensor networks.  We assume
that $\theta_k$  follows a Ornstein-Uhlenbeck  process \cite{karvar12allerton}  
with   correlation      $\cov(\theta_{k_1}, \theta_{k_2}) = \sigma_\theta^2 e^{-| k_1-k_2|/\rho_{\mathrm{corr}}}$ for $k_1 \in [K]$ and $k_2 \in [K]$, where $\rho_{\mathrm{corr}}$ is a parameter that governs the correlation strength,  namely,  a larger  (or  smaller)
$\rho_{\mathrm{corr}}$
corresponds  to  a  weaker  (or  stronger)
correlation.
The covariance matrix of $\boldsymbol \theta$
is  given by
\begin{align*}
\boldsymbol \Sigma_{\boldsymbol \theta} = \sigma_\theta^2
\begin{bmatrix}
1 & e^{- \rho_{\mathrm{corr}}}   & \cdots & e^{-(K-1)\rho_{\mathrm{corr}}} \\
 e^{- \rho_{\mathrm{corr}}} & 1    & \cdots & e^{-(K-2)\rho_{\mathrm{corr}}} \\
\vdots & \vdots   & \ddots & \vdots \\
 e^{- (K-1)\rho_{\mathrm{corr}}} & e^{- (K-2) \rho_{\mathrm{corr}}}   & \cdots &1
\end{bmatrix}.
\end{align*}
where unless specified otherwise, we set $\sigma_\theta^2  = 1$ and $\rho_{\mathrm{corr}} = 0.5$.
The spatial placement and neighborhood structure of the sensor network is modeled by a
random geometric graph \cite{karvar13}, $\mathrm{RGG}(N,d)$, where $N = 10$  sensors
are randomly deployed over a unit square and \textit{bidirectional}
communication links are possible only for pairwise distances at
most $d$. 
Clearly, the topology matrix $\mathbf A$ is determined by $\mathrm{RGG}(N,d)$, and the number of collaboration links increases as $d$ increases. 
In our numerical examples unless specified otherwise,  we set $d = 0.3$ which leads to
 $\mathrm{RGG}(10,0.3)$ shown in Fig.\,\ref{fig: RGG}.

\begin{figure}[htb]
\centerline{ \begin{tabular}{c}
\hspace*{-0.11in}\includegraphics[width=.486\textwidth,height=!]{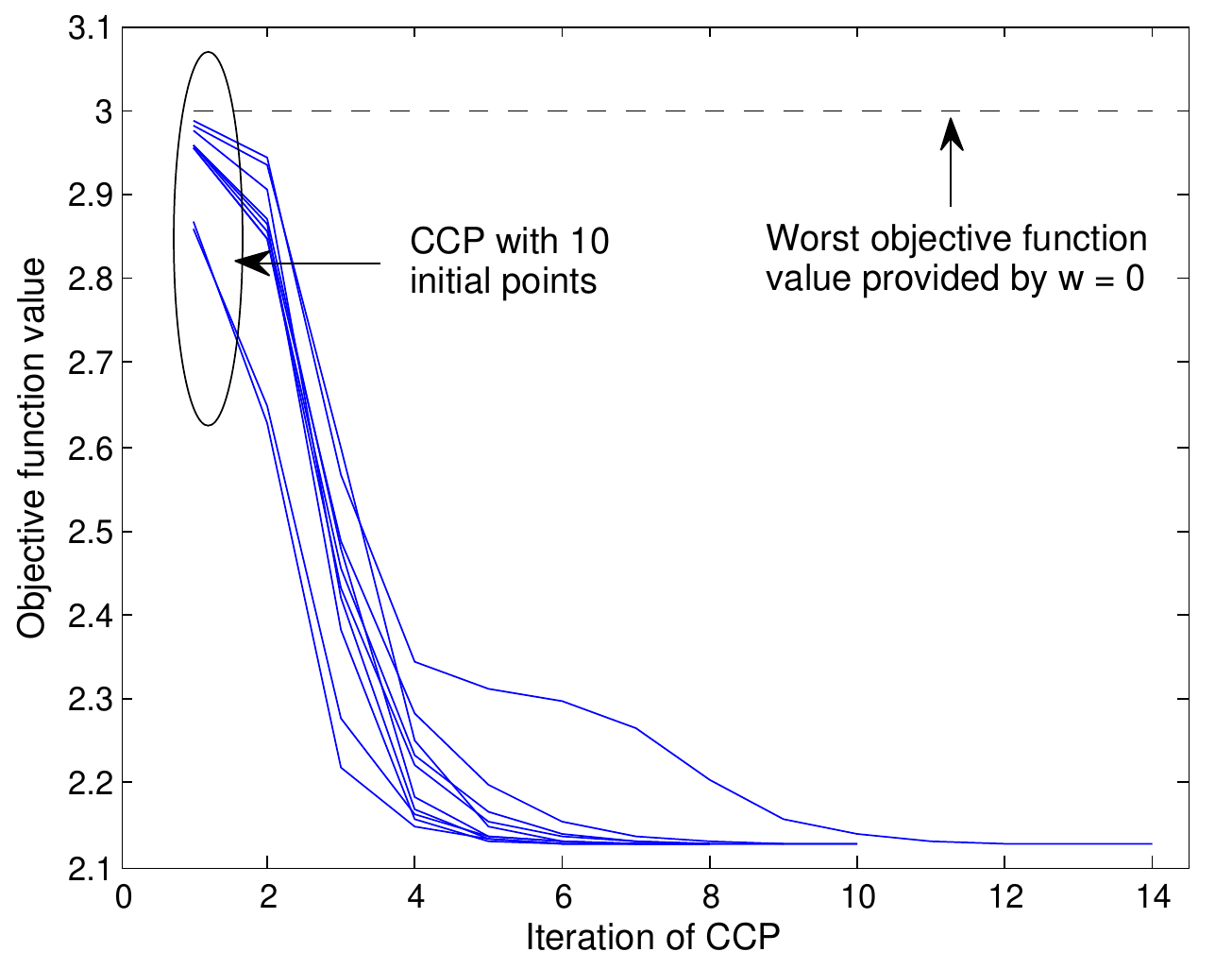}\\
(a) \\
\includegraphics[width=.5\textwidth,height=!]{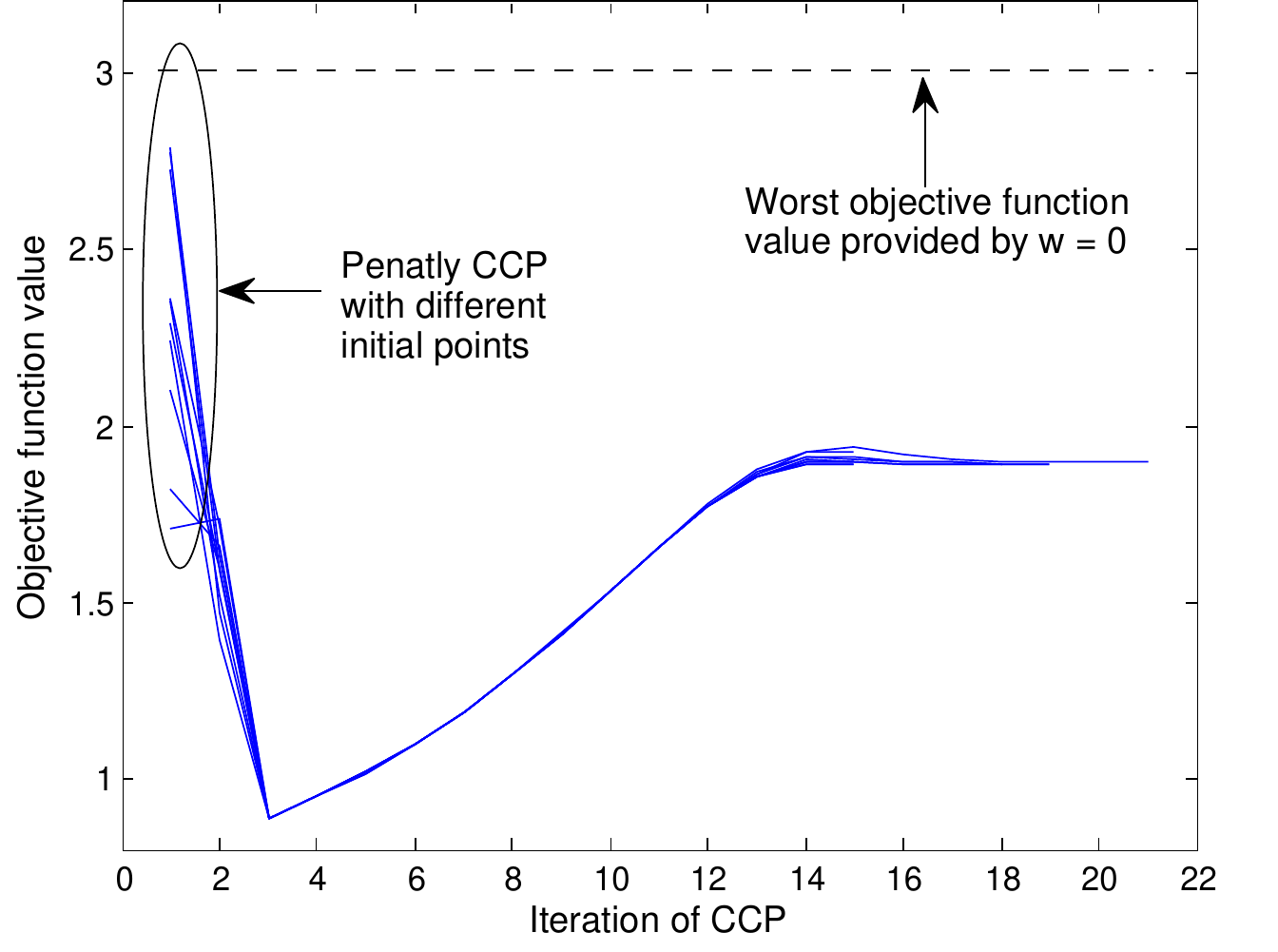}
\\
(b)
\end{tabular}}
\caption{\footnotesize{
Convergence of Algorithm\,1 and 2 for  different initial points.
}}
  \label{fig: converge}
\end{figure}

In the collaborative estimation system shown in Fig.\,\ref{fig: sysmodel}, we  assume that $ M  = N$, $K = 3$,    
$\sigma_{\epsilon}^2 = \sigma_{\varsigma}^2 = 1$,  and
$E_m =   E_{\mathrm{total}}/M$ for $m \in [M]$, where $E_{\mathrm{total}} = 1$ gives the total energy budget of $M$ sensors. For simplicity,  
the   obverstion gain $\mathbf h_k$ and channel gain $\mathbf g_k$ are randomly chosen   from the uniform distribution $\mathcal U(0.1,1)$.  
Moreover, we select  $\tau^0 = 0.1$, $\mu = 1.5$, $\tau_{\max} = 100$ in penalty CCP (namely, Algorithm\,2), 
$a_1 = 0.02$ and $a_2 = 0.5$ in backtracking line search (namely, Algorithm\,4)
and
$\epsilon_{\mathrm{ccp}} = \epsilon_{\mathrm{admm}} = \epsilon_{\mathrm{grad}} = 10^{-3}$ for the stopping tolerance of the proposed algorithms. 
Unless specified otherwise, the ADMM algorithm is adopted at Step\,2 of penalty CCP, and we use CVX \cite{cvx} for all other computations. 
The estimation performance is measured  through  the  empirical  mean squared error (MSE),  which  is  computed  over
$1000$
numerical trials.


In Fig.\,\ref{fig: converge}, we present     convergence  trajectories  of CCP (namely, Algorithm\,1)   and  penalty CCP (namely, Algorithm\,2) as   functions of interation index for $10$ different initial points. 
For comparison, we  plot the worst objective function value of collaboration problem \eqref{eq: prob_main} when $\mathbf w  = \mathbf 0$, namely,   LMMSE is   determined only by the prior information, which leads to the worst  estimation error $\tr(\boldsymbol \Sigma_\theta) = K = 3$.
As we can see, much of the benefit of using
CCP or penalty CCP is gained during the  first few iterations. And each   algorithm  converges to almost the same objective function value for different initial points.
{Compared to CCP, 
the convergence trajectory of penalty CCP is not monotonically decreasing. 
Namely, penalty CCP is not a descent algorithm. The non-monotonicity of penalty CCP is caused by the  penalization on the violation of constraints in the objective function. The objective function value of penalty CCP converges until the penalization ceases to change significantly (after $15$ iterations in this example).}


\begin{figure}[htb]
\centering
\hspace*{-0.1in}\includegraphics[width=.5\textwidth,height=!]{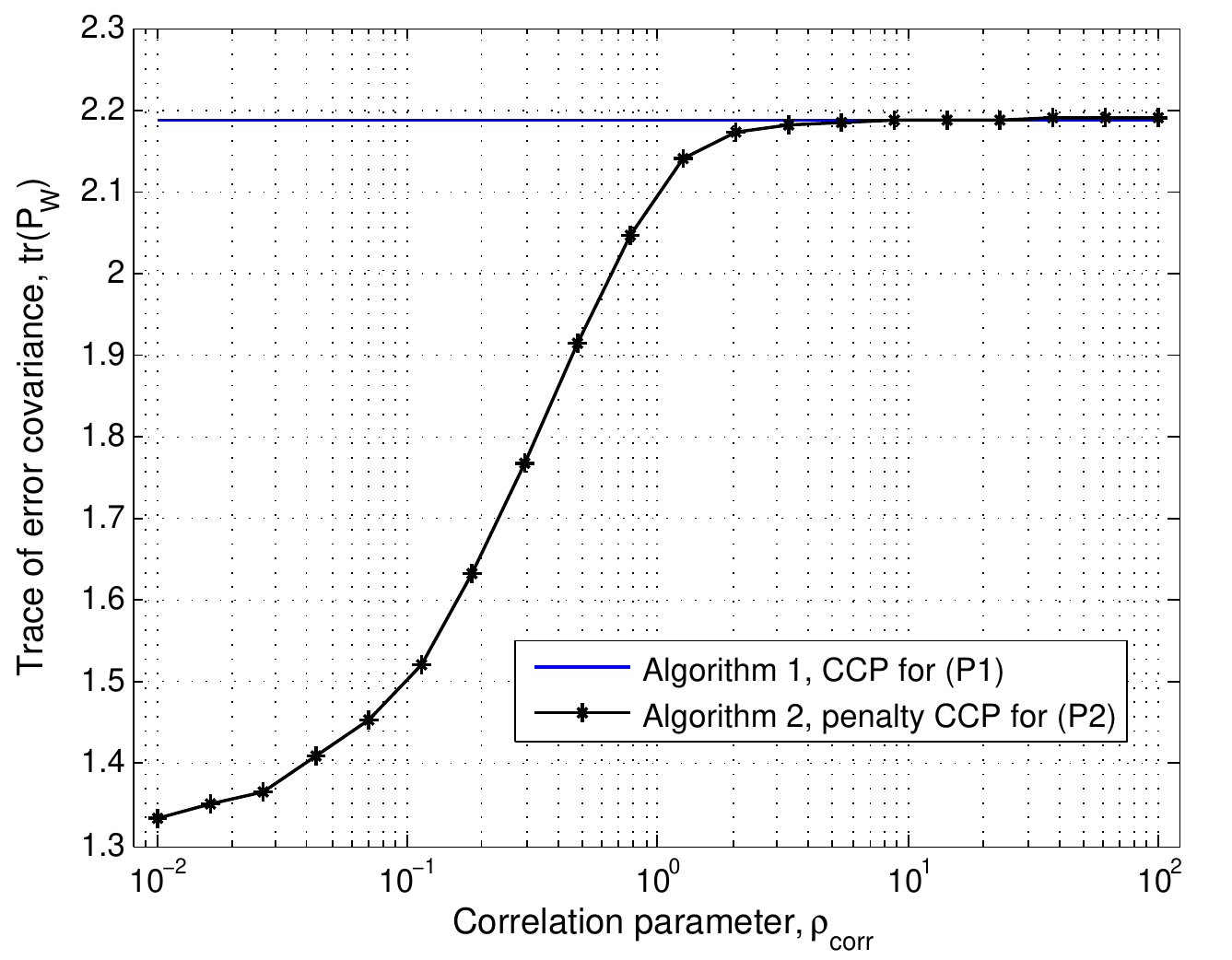} 
\caption{\footnotesize{Estimation error versus correlation parameter $\rho_{\mathrm{corr}}$.
}}
  \label{fig: MSE_corr}
\end{figure}

In Fig.\,\ref{fig: MSE_corr},  we 
 present the trace of error covariance matrix $\mathbf P_W$ given by \eqref{eq: MSE_LMMSE} as a function of the correlation parameter $\rho_{\mathrm{corr}}$, where the sensor collaboration scheme is obtained from 
Algorithm\,1 and Algorithm\,2 to
solve
\eqref{eq: prob_main_uncorr} and \eqref{eq: prob_main_LMMSE}, respectively.
We observe that the estimation error resulting from the solution of 
\eqref{eq: prob_main_uncorr}
 remains unchanged for different values of  $\rho_{\mathrm{corr}}$ since the formulation of \eqref{eq: prob_main_uncorr} is independent of the
prior knowledge about   parameter  correlation. 
The estimation error resulting from the solution of  \eqref{eq: prob_main_LMMSE} increases as $\rho_{\mathrm{corr}}$ increases, and  it eventually converges to the error resulting from the solution of   \eqref{eq: prob_main_uncorr} at an extremely large $\rho_{\mathrm{corr}}$, where parameters become uncorrelated.
This is not surprising, since
 the prior  information about parameter correlation was taken into account in \eqref{eq: prob_main_LMMSE}, thereby significantly  improving the estimation performance.

\begin{figure}[htb]
\centering
\includegraphics[width=.5\textwidth,height=!]{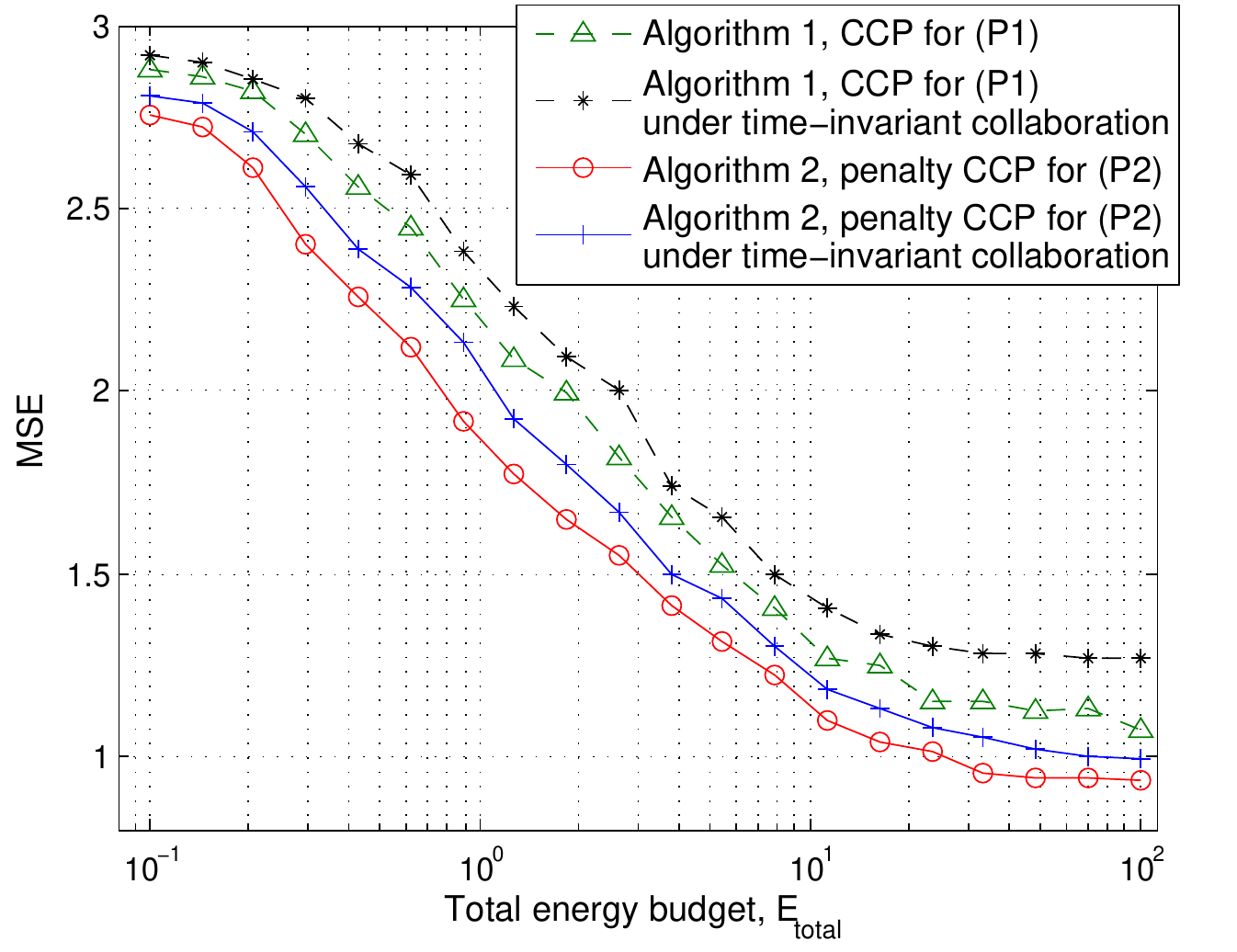} 
\caption{\footnotesize{MSE versus total energy budget.
}}
  \label{fig: MSE_ene}
\end{figure}

In Fig.\,\ref{fig: MSE_ene}, we present the MSE of collaborative estimation as a function of the total energy budget $E_{\mathrm{total}}$ for   $\rho_{\mathrm{corr}} = 0.5$. For comparison, we  plot the estimation performance when using a time-invariant  collaboration scheme 
to solve
\eqref{eq: prob_main_uncorr} and \eqref{eq: prob_main_LMMSE}, respectively. The assumption of time-invariant  collaboration 
implicitly  adds the additional constraint $\mathbf w_1  = \ldots = \mathbf w_K$, which  reduces  the problem size. By fixing the type of algorithm, we observe that 
the MSE  when using  time-invariant sensor collaboration is larger 
than that of the originally proposed  algorithm.
This is because the latter
accounts for temporal dynamics of the network, where  
 observation and channel gains vary in time. Moreover,   the solution of \eqref{eq: prob_main_LMMSE}  yields lower MSE than that of \eqref{eq: prob_main_uncorr}. This result is consistent with Fig.\,\ref{fig: MSE_corr} for a fixed correlation parameter. Lastly, the estimation error is smaller as more energy is used in sensor collaboration.

\begin{figure}[htb]
\centering
\includegraphics[width=.5\textwidth,height=!]{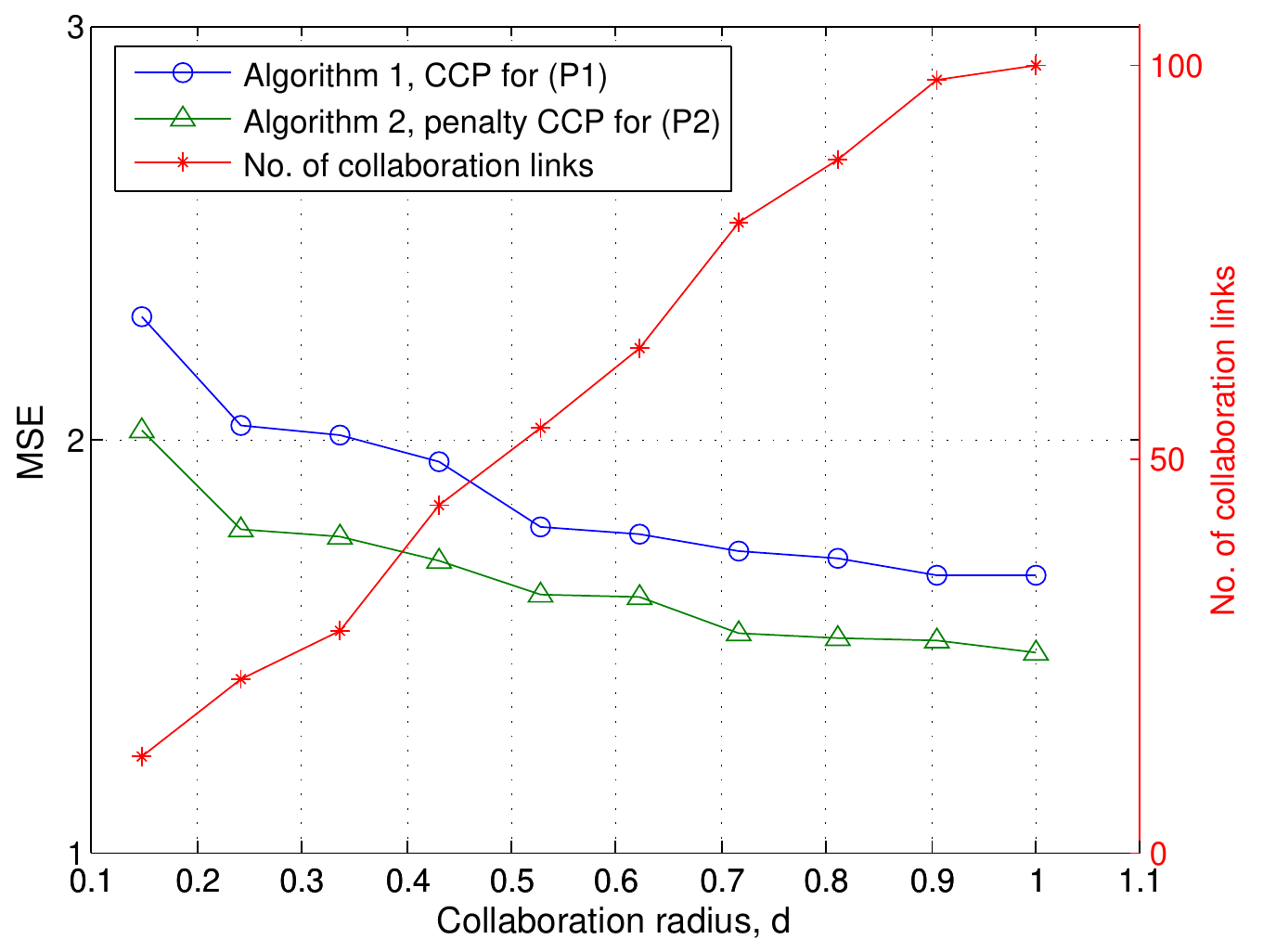} 
\caption{\footnotesize{MSE and collaboration links versus collaboration radius $d$.
}}
  \label{fig: MSE_d}
\end{figure}

In Fig.\,\ref{fig: MSE_d}, we present the MSE and the number of collaboration links as  functions of the collaboration radius $d$ for $\rho_{\mathrm{corr}} = 0.5$ and $E_{\mathrm{total}} = 1$. 
We note that
   the estimation accuracy improves as $d$ increases, since 
  a larger value of $d$ corresponds to  more collaboration links in the network.  
For a fixed value of $d$, the MSE when solving \eqref{eq: prob_main_LMMSE}  is lower than that when solving  \eqref{eq: prob_main_uncorr}, since the latter ignores the information about parameter correlation.  Moreover, we observe that 
the MSE tends to saturate
beyond a collaboration radius $d \approx 0.7$. This indicates that 
  a large part of the performance improvement is achieved only
through partial collaboration. 

{In Fig.\,\ref{fig: noisy_col}, we present   the MSE as a function of the signal-to-noise ratio (SNR), $10\log_{10}( \sigma_{\theta}^2/\sigma_v^2)$, where  $\sigma_{\theta}^2 = 1$ is the variance of the parameter to be estimated, and $\sigma_v^2 \in [10^{-3}, 10^3] $ is the variance of the additive communication noise when  inter-sensor collaboration occurs.  
In this numerical example, 
we study the impact of \textit{noisy} collaboration links  on   estimation performance, where
the  collaboration scheme is obtained  by the solution of \eqref{eq: prob_main_LMMSE} for $d \in \{ 0.5, 1\}$.  As we can see,   estimation distortion increases when SNR decreases. Moreover, the MSE in the presence of     noisy collaboration under the lowest SNR is consistent with that of using the classical amplify-and-forward transmission strategy in the absence of sensor collaboration.
This is because  
each sensor has access to its own measurement in a noiseless manner (collaboration noise only occurs if two different sensors are communicating). 
At a fixed value of SNR, we observe that the MSE decreases as $d$ increases, and it converges to the MSE in the absence of collaboration noise. This implies that the act of sensor collaboration is able to improve   estimation performance even if the collaboration link is noisy.
\begin{figure}[htb]
\centering
\hspace*{-0.1in}\includegraphics[width=.51\textwidth,height=!]{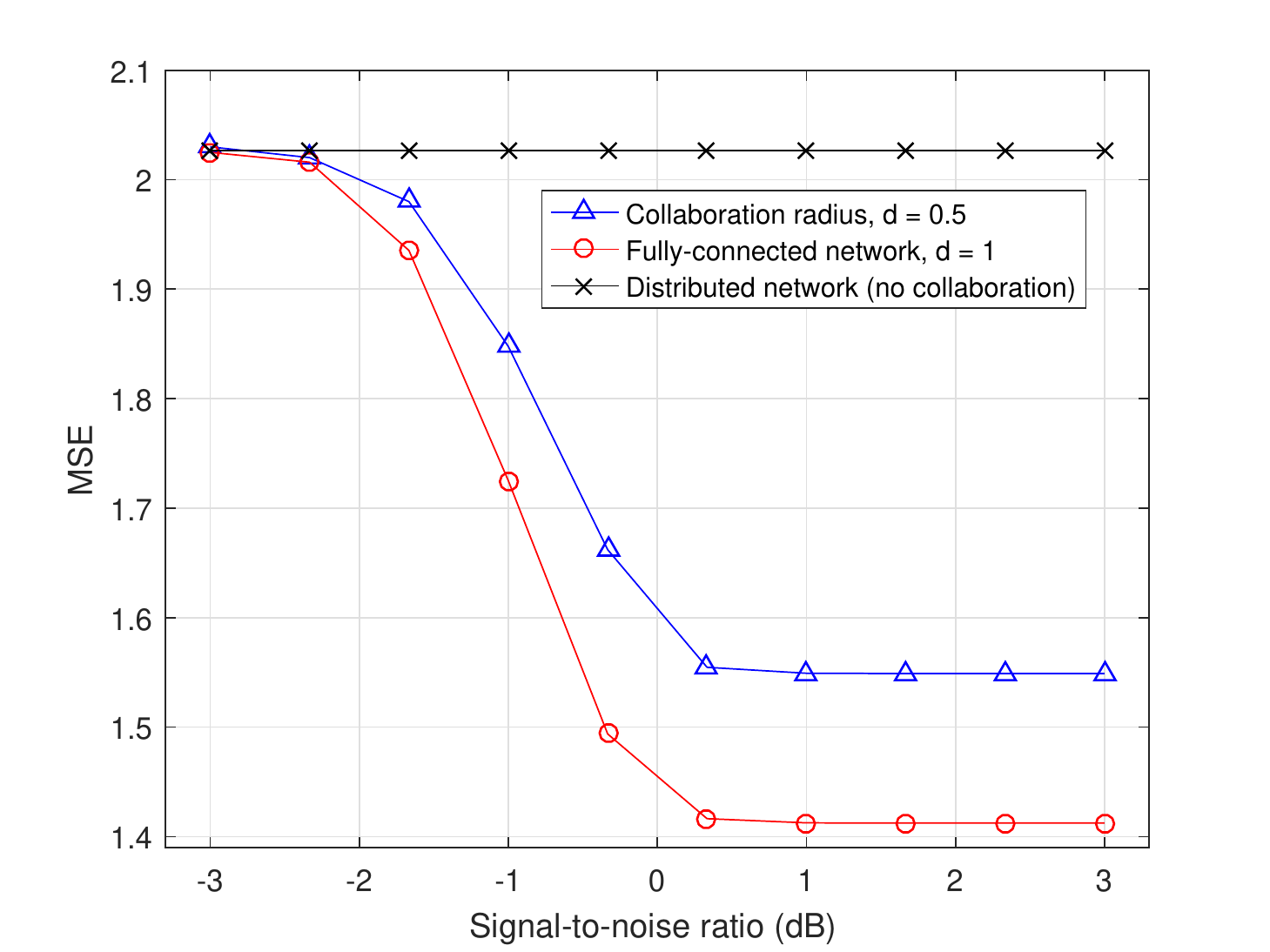} 
\caption{\footnotesize{Noisy collaboration: MSE versus SNR.
}}
  \label{fig: noisy_col}
\end{figure} 
}

\begin{figure}[htb]
\centering
\hspace*{-0.13in}\includegraphics[width=.5\textwidth,height=!]{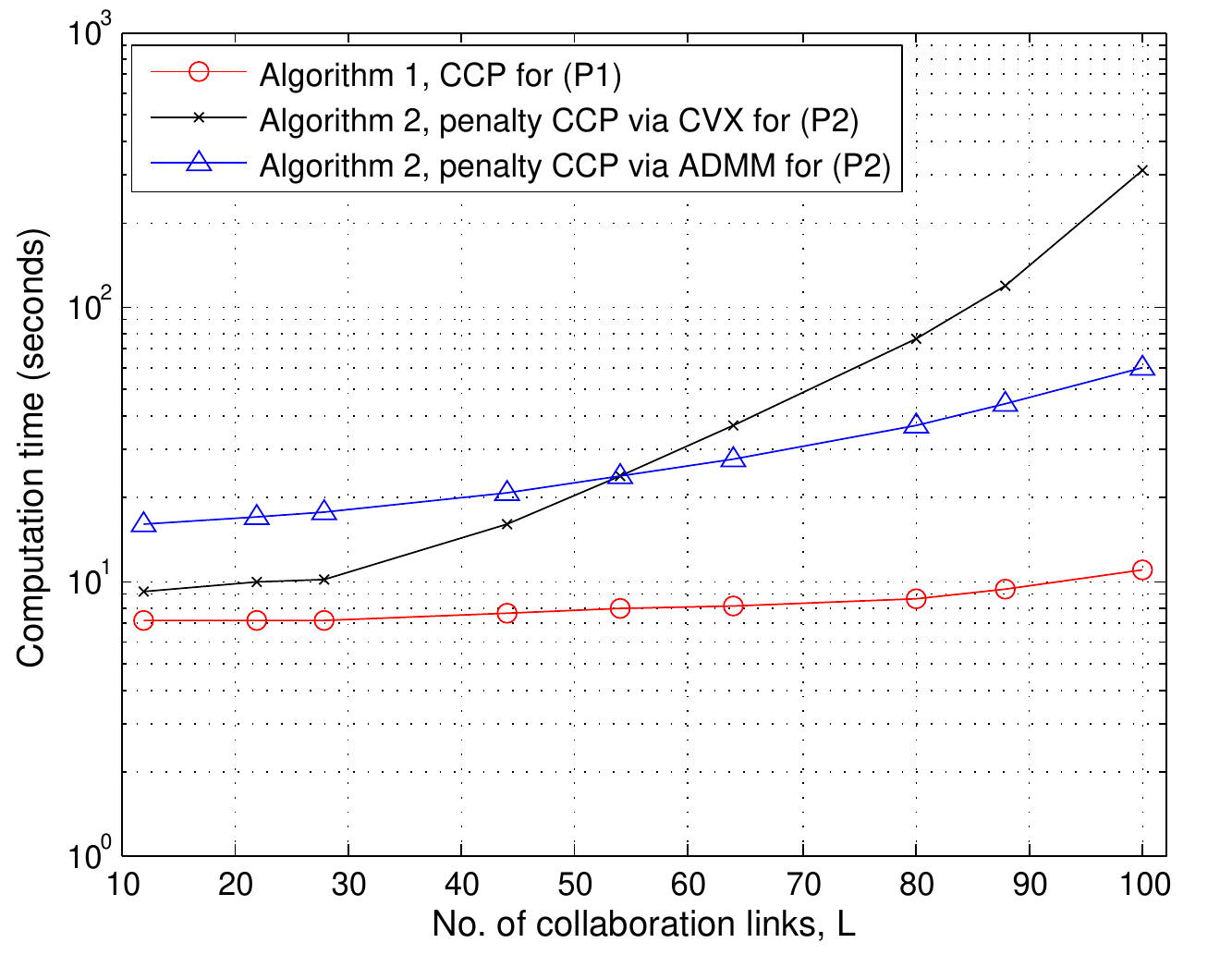} 
\caption{\footnotesize{Computation time versus number of collaboration links.
}}
  \label{fig: computation}
\end{figure}

In Fig.\,\ref{fig: computation}, we present the computation time of our algorithms as   functions of  problem size specified   in terms of the number of collaboration links $L$.
For comparison, we plot the computation time of penalty CCP  when using  an interior-point solver   in CVX \cite{cvx}. 
As we can see, penalty CCP requires  much higher computation time than  CCP, since the former requires   solutions of  SDPs.
When    $L$ is small, we observe that the ADMM based penalty CCP   has a higher computation time than when using  the interior-point solver. This is because  the gradient descent method in ADMM takes relatively more iterations (compared to   small $L$)  to converge with satisfactory accuracy. However,
the ADMM based algorithm performs much faster for a relatively large problem with $L > 80$.

\section{Conclusions}
\label{sec: conc}
We study the problem of sensor collaboration for   estimation of time-varying parameters in   sensor networks. 
Based on
 prior knowledge about  parameter correlations, 
the resulting sensor collaboration problem is solved for estimation of temporally uncorrelated   and correlated parameters. 
In the case of temporally uncorrelated parameters,  
we show that the sensor collaboration problem can be cast as a special nonconvex optimization problem, where a difference of convex functions carries all the nonconvexity. 
By exploiting  problem structure, we solve the problem by using a convex-concave procedure, which renders a good locally optimal solution as evidenced by numerical results.
In the case of   correlated parameters, we show that  the sensor collaboration problem can be converted into a semidefinite program together with a nonconvex rank-one constraint. Spurred by   problem structure, we employ  a semidefinite programming based   penalty convex-concave procedure  to solve the sensor collaboration problem. 
Moreover, we propose an ADMM-based   algorithm that scales more gracefully for large problems.  Numerical results are provided to
demonstrate the effectiveness of our approach   and the impact
of parameter correlation and temporal dynamics of   sensor networks on    the performance of distributed estimation with sensor collaboration.

There are multiple directions for future research. 
{We would like to consider noise-corrupted or quantization-based  imperfect communication links in sensor collaboration. It will  also be of interest to seek the duality gap between the nonconvex sensor collaboration   problems    in order to gain theoretical insights  on the performance of the proposed  optimization methods. 
Another direction of future work
is to seek an approach that  jointly  designs the optimal power allocation scheme  and the collaboration topology. Last but not the  least, it will be worthwhile to study the sensor collaboration problem   in the framework of consensus-based decentralized estimation.
}

\appendices

\section{Proof of Equation \eqref{eq: prop1}
}
\label{appendix: Prop1}{
Let $\mathbf w \in \mathbb R^L$ be the   vector of stacking the \textit{nonzero} entries of $\mathbf W \in \mathbb R^{M \times N}$ columnwise.  
We note that there exists a one-to-one mapping between the element of $\mathbf w$ and the nonzero entry of $\mathbf W$. That is, given $w_{l}$ for  $l \in [L]$, we have a certain pair of indices $( m_l, n_l)$ such that  $w_{l} = W_{m_l n_l}$, where  $m_l \in [M]$ and  $n_l \in [N]$.  Moreover, we obtain that
\begin{align}
W_{ij} = 0, ~ \text{if $(i,j) \notin \mathcal I$ }, 
\end{align}
where $\mathcal I \Def \{ (m_l, n_l) \}_{l=1}^L$.

Given $\mathbf b \in \mathbb R^M$, we have
\[
\mathbf b^T \mathbf W = \begin{bmatrix}
\sum_{i=1}^M b_i W_{i1} &  \cdots & \sum_{i=1}^M b_i W_{iN}
\end{bmatrix}.
\]
Given $\mathbf B \in \mathbb R^{L \times N}$, we obtain 
\[
\mathbf w^T \mathbf B = \begin{bmatrix}
\sum_{l=1}^L B_{l1} W_{m_l n_l} &  \cdots & \sum_{l=1}^L B_{lN} W_{m_l n_l}
\end{bmatrix},
\]
where we used the fact that $w_{l} = W_{m_l n_l}$.

Consider the $t$th entry of $\mathbf w^T \mathbf B$ for $t \in [N]$, we obtain
\begin{align}
[\mathbf w^T \mathbf B ]_t  =& \sum_{l=1}^L B_{lt} W_{m_l n_l} = \sum_{l=1,n_l = t}^L b_{m_l} W_{m_l t} \nonumber \\
=& \sum_{m_l = 1}^M b_{m_l} W_{m_l t}  = [\mathbf b^T \mathbf W ]_t, \label{eq: equivalence}
\end{align}
where we  used the facts  that
$
B_{lt} = \left \{
\begin{array}{ll}
b_{m_l} &  t = n_l \\
0 & \text{otherwise}
\end{array}
\right. 
$
for $(m_l, n_l) \in \mathcal I$,  and $W_{m_l t} = 0$ if $(m_l, t) \notin \mathcal I$.
Based on \eqref{eq: equivalence}, we can conclude that
$ \mathbf w^T \mathbf B = \mathbf b^T \mathbf W$. \hfill $\blacksquare$ 
}

\section{Application of   ADMM}
\label{appendix: ADMM}

We introduce   slack variables   $\boldsymbol \lambda_m \in  \mathbb R^{KL+1}$ for $m \in [M]$ to rewrite \eqref{cons: ene_LMMSE} as an equality constraint together with a second-order cone constraint,
\begin{align}
\bar{\mathbf Q}_m \mathbf w - \boldsymbol\lambda_m + \mathbf c_m = \mathbf 0,
~
\| [\boldsymbol \lambda_m]_{1:KL} \|_2 \leq [\boldsymbol \lambda_m]_{KL+1}, 
\label{eq: SOC_admm}
\end{align}
where $\bar{\mathbf Q}_m  \Def [\mathbf Q_m^{\frac{1}{2}}, \mathbf 0]^T$, 
$ \mathbf Q_m^{\frac{1}{2}}$ is the square root of $\mathbf Q_m$ given by the matrix decomposition
 $\mathbf Q_m =  ( \mathbf Q_m^{\frac{1}{2}} )^T \mathbf Q_m^{\frac{1}{2}}$, 
$ \mathbf c_m = [\mathbf 0^T, \sqrt{E_m}]^T$, and $[\mathbf a]_{1:n}$ denotes a subvector of $\mathbf a$ that consists of its first $n$ entries.

We   further introduce slack variables $\boldsymbol \Lambda_1 \in \mathbb S^{2K}$, $\boldsymbol \Lambda_{2,k} \in \mathbb S^{N+1}$, $\boldsymbol \Lambda_{3,k} \in \mathbb S^{L+1}$ and $\boldsymbol \Lambda_{4,k} \in \mathbb S^{L}$ for $k \in [K]$ to rewrite LMIs of problem \eqref{eq: prob_corr_penalty} as a sequence of equality constraints together with positive semidefinite cone constraints
\begin{align}
&\begin{bmatrix}
\mathbf C -\diag (\mathbf p)   & \mathbf I \\
\mathbf I &  \mathbf V
\end{bmatrix} - \boldsymbol \Lambda_1 = \mathbf 0  \label{eq: LMI1_admm} \\
& \begin{bmatrix} 
p_k & \sigma_{\epsilon}^{-1}  \mathbf h_k^T\\
\sigma_{\epsilon}^{-1}  \mathbf h_k &  \mathbf I +\sigma_{\epsilon}^{2}  \sigma_{\varsigma}^{-2}  \mathbf G_k^T \mathbf U_k \mathbf G_k 
\end{bmatrix}  - \boldsymbol \Lambda_{2,k}= \mathbf 0  \label{eq: LMI2_admm}\\
 & \begin{bmatrix}
\mathbf U_k & \mathbf w_k  \\
\mathbf w_k^T & 1
\end{bmatrix}  -\boldsymbol \Lambda_{3,k} = \mathbf 0 \label{eq: LMI3_admm} \\
&  \mathbf Z_k -  \mathbf U_k + \hat {\mathbf w}_k \mathbf w_k^T + \mathbf w_k \hat {\mathbf w}_k^T - \hat {\mathbf w}_k \hat {\mathbf w}_k^T - \boldsymbol \Lambda_{4,k} = \mathbf 0, \label{eq: LMI4_admm}
\end{align}
where 
$
 \boldsymbol \Lambda_1 \succeq 0$,  $\boldsymbol \Lambda_{2,k} \succeq 0$, $\boldsymbol \Lambda_{3,k} \succeq 0$, and $\boldsymbol \Lambda_{4,k}   \succeq 0$ for $k \in [K]$.

From \eqref{eq: SOC_admm}\,--\,\eqref{eq: LMI4_admm}, problem \eqref{eq: prob_corr_penalty}  becomes
\begin{align}
\begin{array}{ll}
 \minimize
 & \displaystyle \tr (\mathbf V)  + \tau \sum_{k=1}^K \tr(\mathbf Z_k) + \sum_{m=1}^M \mathcal I_0(\boldsymbol \lambda_m)  \vspace*{0.03in} \\
& \displaystyle+ \mathcal I_1 (\boldsymbol \Lambda_1)  + \sum_{i=2}^4 \sum_{k=1}^K   \mathcal I_i(\boldsymbol \Lambda_{i,k} )   \vspace*{0.03in} \\ 
\st  &  \text{equality constraints in \eqref{eq: SOC_admm}\,--\,\eqref{eq: LMI4_admm}}, 
\end{array}
\label{eq: prob_corr_penalty_admm}
\end{align}
where the optimization variables are $\mathbf w$, $\mathbf p$, $\mathbf V$, $\mathbf U_k$, $\mathbf Z_k $, $\boldsymbol \lambda_m$, $\boldsymbol \Lambda_1 $, and $ \{ \boldsymbol \Lambda_{i,k} \}_{i=2,3,4} $   for $m \in [M]$ and $k \in [K]$, and $\mathcal I_i$ is the indicator function specified by
\begin{align}
& \mathcal I_0(\boldsymbol \lambda_{m}) = \left \{ 
\begin{array}{ll}
0, & \text{if $\| [\boldsymbol \lambda_m]_{1:KL} \|_2 \leq [\boldsymbol \lambda_m]_{KL+1}$}\\
\infty & \text{otherwise},
\end{array}
\right. \label{eq: indicator_socp} \\
& \mathcal I_1( \boldsymbol \Lambda_1) = \left \{ 
\begin{array}{ll}
0, & \text{if $ \boldsymbol \Lambda_1 \succeq 0$}\\
\infty & \text{otherwise},
\end{array}
\right. \\
& \mathcal I_i(\boldsymbol \Lambda_{i,k}) = \left \{ 
\begin{array}{ll}
0, & \text{if $\boldsymbol \Lambda_{i,k} \succeq 0$}\\
\infty & \text{otherwise},
\end{array}
\right. ~ i = 2,3,4.
\label{eq: indicator_SDP12}
\end{align}
It is clear from  problem  \eqref{eq: prob_corr_penalty_admm} that the  introduced  indicator functions helps  to isolate the second-order cone   and positive semidefinite cone constraints with respect to slack variables.

Problem  \eqref{eq: prob_corr_penalty_admm}  is now in a form suitable for the application of ADMM. 
The corresponding augmented Lagrangian  \cite{boyparchupeleck11} in ADMM is given by 
\begin{align}
\label{eq: Lag}
&\mathcal L_\rho (\mathscr X,\mathscr Z,\mathscr Y) = 
\tr (\mathbf V)  + \tau \sum_{k=1}^K \tr(\mathbf Z_k) + \sum_{m=1}^M \mathcal I_0(\boldsymbol \lambda_m)  \nonumber \\
&+ \mathcal I_1 (\boldsymbol \Lambda_1)  + \sum_{i=2}^4 \sum_{k=1}^K   \mathcal I_i(\boldsymbol \Lambda_{i,k} ) + \sum_{m=1}^M \boldsymbol \pi_{m}^T \mathbf f_m (\mathscr X,\mathscr Z) \nonumber \\
&  + \frac{\rho}{2} \sum_{m=1}^M  \|  \mathbf f_m (\mathscr X,\mathscr Z) \|_2^2 + \tr\left (
\boldsymbol \Pi_1^T \mathbf F_1 (\mathscr X,\mathscr Z) 
\right ) \nonumber \\
& + \frac{\rho}{2} \| \mathbf F_1 (\mathscr X,\mathscr Z)  \|_F^2 + \sum_{i=2}^4 \sum_{k=1}^K \tr\left (
\boldsymbol \Pi_{i,k}^T \mathbf F_{i,k} (\mathscr X,\mathscr Z) 
\right ) \nonumber \\
& +  \frac{\rho}{2}  \sum_{i=2}^4 \sum_{k=1}^K \|\mathbf F_{i,k} (\mathscr X,\mathscr Z) \|_F^2,
\end{align}
where 
  $ \mathscr X $ denotes  the  set   of  {primal  variables}  
$\mathbf w$, $\mathbf p$, $\mathbf V$, $\mathbf U_k$ and $\mathbf Z_k $ for $k \in [K]$,  $ \mathscr Z$ denotes   the  set  of  {primal slack   variables}  
$\boldsymbol \lambda_m$, $\boldsymbol \Lambda_1 $ and $ \{ \boldsymbol \Lambda_{i,k} \}_{i=2,3,4} $   for $m \in [M]$ and $k \in [K]$,   $\mathscr Y$ is the set of dual variables  
$\boldsymbol \pi_m$, $\boldsymbol \Pi_1$ and $\{ \boldsymbol \Pi_{i,k} \}_{i=2,3,4}$ for $m \in [M]$  and $k \in [K]$,    
$\mathbf   f_m (\cdot)$, $ \mathbf F_1 (\cdot)$, and  $ \mathbf  F_{i,k}   (\cdot)$    for $i \in \{ 2,3,4\}$ represent   linear functions at the left hand side of   equality constraints in \eqref{eq: SOC_admm}\,--\,\eqref{eq: LMI4_admm},
$\rho >0$ is a regularization parameter, and   $\| \cdot \|_F$  denotes the Frobenius norm of a matrix.


We 
  iteratively execute the following three steps for ADMM iteration $t = 0,1,\ldots$  
\begin{align}
& \mathscr X^{t+1} = \displaystyle \argmin_{\mathscr X} \mathcal L(\mathscr X, \mathscr Z^t, \mathscr Y^t) \label{eq: p_step} \\ 
& \mathscr Z^{t+1} = \displaystyle \argmin_{\mathscr Z} \mathcal L(\mathscr X^{t+1} , \mathscr Z, \mathscr Y^t) \label{eq: s_step} \\ 
& \left \{
\begin{array}{l}
\boldsymbol \pi_{m}^{t+1 } = \boldsymbol \pi_{m}^{t }  + \rho\, \mathbf  f_{m} (\mathscr X^{t+1} , \mathscr Z^{t+1}  ) ,~ \forall m \\
\boldsymbol \Pi_1^{t+1} = \boldsymbol \Pi_1^{t} +  \rho\,  \mathbf F_1 (\mathscr X^{t+1} , \mathscr Z^{t+1}   )   \\
\boldsymbol \Pi_{i,k}^{t+1}  = \boldsymbol \Pi_{i,k}^{t}  + \rho\, \mathbf  F_{i,k}  (\mathscr X^{t+1} , \mathscr Z^{t+1}   ) , ~\forall i, k, 
\end{array} 
\right.
\label{eq: admm_3steps}
\end{align}
until both of the conditions  $\| \mathscr X^{t+1} - \mathscr Z^{t} \|_F  \leq \epsilon_{\mathrm{admm}}    $  and $\|  \mathscr Z^{t+1} - \mathscr Z^{t } \|_F  \leq \epsilon_{\mathrm{admm}}    $  are satisfied, 
where with an abuse of notation, $\| \mathscr X  \|_F $ denotes the sum of Frobenius norms of    variables in  $\mathscr X$, and
$\epsilon_{\mathrm{admm}}$ is a  stopping tolerance.

Substituting  \eqref{eq: Lag} into \eqref{eq: p_step} and completing the squares with respect to   primal variables, 
the $\mathscr X$-minimization problem \eqref{eq: p_step} becomes the unconstrained quadratic program given by \eqref{eq: UQP}.

Substituting  \eqref{eq: Lag} into \eqref{eq: s_step}, 
the $\mathscr Z$-minimization problem \eqref{eq: s_step} is decomposed into a sequence of subproblems
with respect to each of slack variables, given by \eqref{eq: prob_socp_m}, \eqref{eq: prob_SDP_1} and \eqref{eq: prob_SDP_i}.  \hfill $\blacksquare$

\section{Proof of Proposition\,\ref{prop: gradient}}
\label{app: gradient}
We begin by collecting      terms in  $\varphi$ associated with $\mathbf w$,  
\begin{align}
\varphi_{\mathbf w} \Def &  \frac{\rho}{2} \sum_{m = 1}^M \left \| \bar {\mathbf Q}_m \mathbf w - \boldsymbol \alpha_m  \right \|_2^2 +  \rho \sum_{k=1}^K  \| 
\mathbf w_k -  \boldsymbol \gamma_{3,k} 
\|_2^2 \nonumber \\
&+\frac{\rho}{2}  \sum_{k=1}^K  \|  \hat {\mathbf w}_k \mathbf w_k^T + \mathbf w_k \hat {\mathbf w}_k^T - \mathbf H_k  \|_F^2, \label{eq: obj_w}
\end{align}
where  
$\boldsymbol \gamma_{3,k} $ is the $(L+1)$ column of  $\boldsymbol \Upsilon_{3,k}$ after the last entry is removed, and $\mathbf H_k \Def \mathbf U_k - \mathbf Z_k +  \hat {\mathbf w}_k \hat {\mathbf w}_k^T + \boldsymbol \Upsilon_{4,k} $, which is a symmetric matrix.   

In \eqref{eq: obj_w}, we assume an incremental change $\delta \mathbf w$ in $\mathbf w$. Replacing  $\mathbf w$ with $\mathbf w + \delta \mathbf w$ and $\varphi_{\mathbf w} $ with $\varphi_{\mathbf w} + \delta \varphi_{\mathbf w} $
and collecting first order variation terms on both sides of \eqref{eq: obj_w}, we obtain
\begin{align}
\delta \varphi_{\mathbf w} =  &   \rho \sum_{m=1}^M 
(\bar {\mathbf Q}_m \mathbf w - \boldsymbol \alpha_m )^T \bar {\mathbf Q}_m   \delta \mathbf w   + 2 \rho (\mathbf w - \boldsymbol \gamma_3)^T \delta \mathbf w  \nonumber \\
& +2 \rho \hat{\mathbf w}^T \mathrm{blkdiag}\{\hat {\mathbf w}_k \mathbf w_k^T + \mathbf w_k \hat {\mathbf w}_k^T - \mathbf H_k \} \delta \mathbf w,
\label{eq: delta_w}
\end{align}
where $\boldsymbol \gamma_3 = [ \boldsymbol \gamma_{3,1}^T, \ldots, \boldsymbol \gamma_{3,K}^T]^T$, and $\hat{\mathbf w} = [\hat{\mathbf w}_1^T, \ldots, \hat{\mathbf w}_K^T]^T$. It is clear from \eqref{eq: delta_w} that the gradient of $\varphi$  with respect to  $\mathbf w$ is given by
\begin{align}
\nabla_{\mathbf w} \varphi =   &\rho \sum_{m=1}^M \bar {\mathbf Q}_m^T  
(\bar {\mathbf Q}_m \mathbf w - \boldsymbol \alpha_m ) +  2 \rho (\mathbf w - \boldsymbol \gamma_3) \nonumber \\
&+ 2 \rho \,  \mathrm{blkdiag}\{\hat {\mathbf w}_k \mathbf w_k^T + \mathbf w_k \hat {\mathbf w}_k^T - \mathbf H_k \} \hat{\mathbf w}. \label{eq: grad_w}
\end{align}

Second, we collect the terms associated with $\mathbf p$   in  $\varphi$ to construct the function
\begin{align}
\varphi_{\mathbf p} \Def \frac{\rho}{2} \| \mathbf C - \mathrm{diag}(\mathbf p) - \boldsymbol \Upsilon_1^{11} \|_F^2 + \frac{\rho}{2} \| \mathbf p - \boldsymbol \gamma_2 \|_2^2,\label{eq: obj_p}
\end{align}
where
$\boldsymbol \Upsilon_1^{11}$ is a matrix that consists    of  the first $K$  rows and columns of $\boldsymbol \Upsilon_1$,  and $\boldsymbol \gamma_2 $ is a vector whose $k$th entry is given by the first entry of $\boldsymbol \Upsilon_{2,k}$ for $k \in [K]$. 

In \eqref{eq: obj_p}, replacing  $\mathbf p$ with $\mathbf p + \delta \mathbf p$ and $\varphi_{\mathbf p} $ with $\varphi_{\mathbf p} + \delta \varphi_{\mathbf p} $ and collecting first order variation terms on both sides, we obtain
\begin{align}
\delta \varphi_{\mathbf p} = \rho [  2 \mathbf p + \diag(\boldsymbol \Upsilon_1^{11} )  -\diag(\mathbf C)  - \boldsymbol \gamma_2]^T \delta \mathbf p,
\end{align}
where $\diag(\cdot)$ returns in vector form the diagonal entries of its matrix argument.
Therefore, the gradient of $\varphi$ with respect to $\mathbf p$ is given by
\begin{align}
\nabla_{\mathbf p} \varphi =   \rho [  2 \mathbf p + \diag(\boldsymbol \Upsilon_1^{11} )  -\diag(\mathbf C)  - \boldsymbol \gamma_2]. \label{eq: grad_p}
\end{align}

Third, 
given the terms associated with $\mathbf V$   in  $\varphi$, the gradient of $\varphi$ with respect to $\mathbf V$ is readily cast as
\begin{align}
\nabla_{\mathbf V} \varphi =  \mathbf I + \rho(\mathbf V - \boldsymbol \Upsilon_1^{22}), \label{eq: grad_V}
\end{align}
where $\boldsymbol \Upsilon_1^{22}$ is a submatrix of  $\boldsymbol \Upsilon_1$ after the first $K$ rows and columns are removed. 

Further, we collect the terms in   $\varphi$ with respect to the variable $\mathbf U_k$, and consider the function
\begin{align}
\varphi_{\mathbf U_k} \Def& ~ \frac{\rho}{2}  \left \|  \mathbf I +\sigma_{\epsilon}^{2}  \sigma_{\varsigma}^{-2}  \mathbf G_k^T \mathbf U_k \mathbf G_k  - \boldsymbol \Upsilon_{2,k}^{22} \right \|_F^2 \nonumber \\
&+ \frac{\rho}{2} \| 
\mathbf U_k - \boldsymbol \Upsilon_{3,k}^{11}
\|_F^2
+  \frac{\rho}{2} \| 
\mathbf U_k - \mathbf Z_k - \mathbf T_k
\|_F^2, \label{eq: obj_Uk}
\end{align}
where $\boldsymbol \Upsilon_{2,k}^{22} $ is   a submatrix of  $\boldsymbol \Upsilon_{2,k}$ after the first  row and column  are removed, $\boldsymbol \Upsilon_{3,k}^{11}$ is a   submatrix of  $\boldsymbol \Upsilon_{3,k}$ after the last  row and column  are removed, and $\mathbf T_k \Def   \hat {\mathbf w}_k \mathbf w_k^T + \mathbf w_k \hat {\mathbf w}_k^T - \hat {\mathbf w}_k \hat {\mathbf w}_k^T - \boldsymbol \Upsilon_{4,k} $.

In \eqref{eq: obj_Uk}, replacing  $\mathbf U_k$ with $\mathbf U_k+ \delta \mathbf U_k$ and $\varphi_{\mathbf U_k} $ with $\varphi_{\mathbf U_k} + \delta \varphi_{\mathbf U_k} $ and collecting first order variation terms on both sides, we obtain
\begin{align*}
  &\delta \varphi_{\mathbf U_k}     =   
\frac{ \rho\sigma_{\epsilon}^{2} }{\sigma_{\varsigma}^{2}  }
\tr \left ( \mathbf G_k
(\mathbf I +\frac{  \sigma_{\epsilon}^{2} }{\sigma_{\varsigma}^{2}  } \mathbf G_k^T \mathbf U_k \mathbf G_k  - \boldsymbol \Upsilon_{2,k}^{22})^T  \mathbf G_k^T \delta \mathbf U_k 
\right) \nonumber \\
&  \quad + \rho \tr\left (
\left (\mathbf U_k - \boldsymbol \Upsilon_{3,k}^{11})^T \delta \mathbf U_k  +
(\mathbf U_k  - \mathbf Z_k - \mathbf T_k \right )^T \delta \mathbf U_k 
\right).
\end{align*}
Therefore, the gradient of $\varphi$ with respect to $\mathbf U_k$ is given by
\begin{align}
\nabla_{\mathbf U_k} \varphi = & { \rho\sigma_{\epsilon}^{2} }{\sigma_{\varsigma}^{-2}  } \mathbf G_k
(\mathbf I + {  \sigma_{\epsilon}^{2} }{\sigma_{\varsigma}^{-2}  } \mathbf G_k^T \mathbf U_k \mathbf G_k  - \boldsymbol \Upsilon_{2,k}^{22})   \mathbf G_k^T \nonumber \\
& + \rho (\mathbf U_k - \boldsymbol \Upsilon_{3,k}^{11}) + \rho (\mathbf U_k -\mathbf Z_k - \mathbf T_k). \label{eq: grad_Uk}
\end{align}

Finally, the gradient of $\varphi$ with respect to $\mathbf Z_k$ is given by 
\begin{align}
\nabla_{\mathbf Z_k} \varphi =  \tau \mathbf I + \rho (\mathbf Z_k - \mathbf U_k +\mathbf T_k ), \label{eq: grad_Zk}
\end{align}
where $\mathbf T_k$ is defined in \eqref{eq: obj_Uk}.
We now complete the proof  by combining  \eqref{eq: grad_w}, \eqref{eq: grad_p}, \eqref{eq: grad_V}, \eqref{eq: grad_Uk} and \eqref{eq: grad_Zk}.
\hfill $\blacksquare$

\bibliographystyle{IEEEbib}
\bibliography{journal_sel,journal_col,journal_EH}

\end{document}